\documentclass[onecolumn]{aastex63}
\usepackage{graphicx}

\def\simlt{\lower.5ex\hbox{$\; \buildrel < \over \sim \;$}}
\def\simgt{\lower.5ex\hbox{$\; \buildrel > \over \sim \;$}}

\def\cm3{{\rm cm^{-3}}}
\def\pcc{{\rm\,cm^{-3}}}
\def\kms{km s$^{-1}$}
\def\Msun{{$M_\odot$}}
\def\Lsun{{$L_\odot$}}

\def\vexp{v_{\rm exp}}
\def\vlsr{v_{\rm LSR}}

\def\g54{G54.1+0.3}

\def\nbg{n_{\rm bg}}

\def\fe164{[Fe II] 1.644~$\mu$m}

\def\h212{H$_2$ 2.122~$\mu$m}
\def\schi{{\sc Hi}}
\def\schii{{\sc Hii}\ }

\def\e{$^{-1}$}

\def\esn{E_{\rm SN}}
\def\esnth{E_{\rm SN}(E_{\rm th})}
\def\esnksh{E_{\rm SN}(E_{\rm K,\,sh})}
\def\esnpsh{E_{\rm SN}(p_{\rm sh})} 

\def\eth{E_{\rm th}}
\def\vshell{v_{\rm sh}}
\def\rshell{R_{\rm sh}}
\def\mshell{M_{\rm sh}}
\def\pshell{p_{\rm sh}}

\def\nbarc{\bar n_{\rm bg}}
\def\ekshell{E_{\rm K, sh}}
\def\tshell{t_{\rm sh}}

\def\tsf{t_{\rm sf}}
\def\rsf{r_{\rm sf}}
\def\psf{p_{\rm sf}}

\def\fdg{.\!\!^\circ}
\def\erg{\,{\rm erg}}
\def\Myr{\,{\rm Myr}}

\def\vexp{v_{\rm exp}}
\def\vlos{v_{\rm LOS}}
\def\rsvol{R_{s,{\rm vol}}}

\def\farcm{\hbox{$.\mkern-4mu^\prime$}}
\def\farcs{\hbox{$.\mkern-4mu^{\prime\prime}$}}

\def\la{\mathrel{\hbox{\rlap{\hbox{\lower4pt\hbox{$\sim$}}}\hbox{$<$}}}}
\def\ga{\mathrel{\hbox{\rlap{\hbox{\lower4pt\hbox{$\sim$}}}\hbox{$>$}}}}

\submitjournal{ApJ}
	
\shorttitle{Radiative Supernova Remnants and Supernova Feedback}
\shortauthors{Koo et al.}

\begin{document}
\title{Radiative Supernova Remnants and Supernova Feedback}

\correspondingauthor{Bon-Chul Koo}
\email{koo@astro.snu.ac.kr}

\author{Bon-Chul Koo}
\affiliation{Department of Physics and Astronomy, Seoul National University \\
Seoul 08826, Korea}

\author{Chang-Goo Kim}

\affil{Department of Astrophysical Sciences, Princeton University, Princeton, NJ 08544, USA}
\affil{Center for Computational Astrophysics, Flatiron Institute, New York, NY 10010, USA}

\email{cgkim@astro.princeton.edu}

\author{Sangwook Park}

\affil{Department of Physics, University of Texas at Arlington, Arlington, TX 76019, USA}
\email{s.park@uta.edu}

\author{Eve C. Ostriker}

\affil{Department of Astrophysical Sciences, Princeton University, Princeton, NJ 08544, USA}
\email{eco@princeton.edu}

\begin{abstract}

Supernova (SN) explosions are 
a major feedback mechanism regulating star formation in galaxies through their momentum input. 
We review the observations of SNRs in radiative stages in the Milky Way to 
validate the theoretical results on the momentum/energy injection from a single SN explosion. 
For seven SNRs where we can observe fast-expanding, atomic radiative  shells, we show that the shell momentum inferred from \schi\ 21 cm line observations 
is in the range of (0.5--4.5)$\times 10^5$ \Msun\ \kms. 
In two SNRs (W44 and IC 443), shocked molecular gas with momentum 
comparable to  that of the atomic SNR shells has been also observed. 
We compare the momentum and kinetic/thermal energy of these seven SNRs 
with the results from 1D and 3D numerical simulations. 
The observation-based momentum and kinetic energy  
agree well with the expected momentum/energy input from 
an SN explosion of $\sim 10^{51}$~erg. 
It is much more difficult to use data/model comparisons of thermal energy 
to constrain the initial explosion energy, however,  
due to rapid cooling and complex physics at the hot/cool interface in radiative SNRs.
We discuss the observational and theoretical uncertainties 
of these global parameters and explosion energy estimates
for SNRs in complex environments.

\end{abstract}


\section{Introduction}\label{sec:intro}

Massive stars have profound impacts on the evolution of the surrounding
interstellar medium (ISM), star formation rates, and baryonic cycles in
galaxies.  The momentum and energy injected by feedback from massive young
stars -- supernovae (SNe), radiation, and stellar winds -- drive turbulence and
heat the gas to regulate star formation rates \citep[e.g.,][]{mckee07,krumholz14}.
Stellar population synthesis models imply that 
the momentum input rates
of these feedback processes averaged with stellar initial mass function are more or less
similar \citep[e.g.,][]{leitherer99,agertz13}. 
However, according to these stellar population synthesis models, 
the duration of SNe feedback is much longer than those of the other feedback mechanisms 
(i.e., $\sim 40\Myr$ vs $< 5\Myr$ after the birth of stellar population), 
which makes SNe the most important dynamical feedback mechanism.
Furthermore, the energy injection of SNe is highly localized in time and space,
creating strong forward and reverse shocks and heating the interior to very
high temperature. The subsequent expansion into the ambient medium 
is nearly adiabatic as long as the shock speed and hence the postshock 
temperature is high ($T>10^6$~K). During the adiabatic stage of evolution, 
the total injected radial momentum is
boosted by more than an order of magnitude \citep{kim15}.
Understanding the momentum boosting process and the level of the terminal
momentum as a consequence of SNR evolution has practical importance in galaxy
formation simulations. In low-resolution simulations, a simple implementation of SN feedback in
the form of a thermal energy dump results in immediate loss of energy (a well
known overcooling problem; e.g., \citealt{katz92}) due to the inability of resolving
the shell formation (or cooling) radius (or mass) \citep{kim15}. 
Since after reaching a  terminal level, the momentum is conserved and not radiated away, injecting
the ``proper'' terminal momentum would be preferred 
to capture the dynamical impact of SN feedback \citep[e.g.,][but see
\citealt{kim18,hu19} for a  discussion of the other important element of SN feedback, hot gas
creation, in galactic scale wind driving]{kimm14,hopkins14,rosdahl17,hopkins18,smith18}.

The momentum boost occurs during the energy conserving stage (also known as the
Sedov-Taylor stage; see \citealt{sedov59,taylor50,draine11}). As the SN blast wave sweeps
up more and more volume and mass ($R\propto t^{2/5}$ and $M\propto t^{6/5}$)
while the energy is conserved, the total radial
momentum increases with time 
$p\propto(ME)^{1/2}\propto t^{3/5}$.
When the temperature of shocked gas falls below $\sim10^6$ K as expansion slows
down, radiative cooling from metal lines (e.g., highly ionized C, O, and N; see
\citealt{sutherland93,gnat12}) becomes strong and a cold, thin shell forms.
Interestingly, since the end of the energy conserving stage (or shell
formation) is determined by shock temperature (or expansion velocity), the
radial momentum ($p\sim E/v$) at the time of shell formation is nearly independent to
the background medium density although the shell formation time itself depends
quite sensitively on the mean density of the ISM \citep[e.g.,][]{draine11,kim15}.
The evolution of SNRs in a uniform medium is well understood
theoretically from both analytic theories
\citep[e.g.,][]{cox72,mckee77,draine11} and direct numerical simulations with
radiative cooling
\citep[e.g.,][]{chevalier74, cioffi88,thornton98,blondin98}.

In recent years, it has become possible to directly simulate SNR evolution in
full three dimensions with high resolution. This allows realistic modeling of
the background state of the ISM where the SN explodes, including the high degree of
inhomogeneity that exists in the real ISM.  Several recent numerical
simulations have been conducted to include the effect of inhomogeneous
background states on SNR evolution. In \citet[][KO15 hereafter]{kim15}, a SN
explodes in a cloudy two-phase ISM (cold clouds embedded in volume-filling warm
diffuse medium) developed from nonlinear saturation of thermal instability.
\citet{li15} placed a SN in a classical three-phase ISM of \citet{mckee77},
with clouds consisting of cold cores and warm envelopes embedded in
a volume-filling hot diffuse medium. 
\citet{zhang19} simulated SNR evolution in a medium realized by driven turbulence. 
To study SNR expansion within molecular clouds, either an imposed log-normal density
distribution \citep{martizzi15, walch15} or clouds with driven turbulence
\citep{iffrig15} has been adopted as a background medium. 
According to these numerical studies, the momentum of the shell at the time of 
formation as well as the terminal momentum is insensitive to the background
density and hence density inhomogeneity. The resulting final momentum injected to the ISM
by a single SN explosion is $p_{\rm SN}=(1-5)\times10^5$~\Msun\kms\ for the
canonical explosion energy $\esn=10^{51}{\rm\,erg}$. 

Although there is an emerging consensus in theoretical work on the SNR
evolution and the final momentum injected to the ISM from a single SN event,
observational constraints are still lacking and necessary to validate the
theoretical understanding. In addition, 
there is a missing physical element, cosmic rays, in most studies of SNR evolution, which
may or may not have a significant impact (of an order of a few) on the terminal
momentum \citep[see][but numerical results are sensitive to the model of cosmic
ray injection, M. Li et al. in prep]{diesing18}. Direct observational constraints on the
SNR momentum and energy will play a crucial role to refine theories and identify missing
physics in contemporary numerical simulations.

Observationally, it is very difficult to detect  
radiative shells associated with SNRs.
There are about 300 known SNRs in the Milky Way, most of which 
are identified in radio continuum \citep{green19}. 
Considering that SNRs spend most of their lifetime in the radiative stage, 
we expect to see fast expanding shells of atomic hydrogen associated with 
many of these SNRs. However, because the SNRs are in the Galactic plane, 
the observation of radiative shells in \schi\ 21 cm line is severely hampered by 
the contamination due to the background/foreground \schi\ emission \citep[e.g.,][]{koo04a}.
There have been systematic searches for radiative \schi\ shells
associated with the {\em known} Galactic SNRs in the northern and southern skies, but the expected fast-expanding \schi\ shells have been detected only towards a 
handful of SNRs \citep{koo91, koo04b}. 
Another approach to identify radiative \schi\ shells associated with SNRs 
is to search for localized high-velocity (HV) \schi\ features in 
{\em unbiased} \schi\ surveys of the Galactic plane.  
The basic idea is that there might be many old, radio-faint SNRs not identified in 
radio continuum but still possess a fast-expanding \schi\ shell. The continuing discovery of 
radio-faint SNRs supports this idea \citep{foster13, gerbrandt14,driessen18,hurley19}. There are a few 
old radiative SNR candidates identified in this approach \citep{koo04a,kang07,kang14}. 

In this paper, we inventory the global parameters 
of radiative SNRs with fast-expanding \schi\ shells 
and compare them with the results of 1D and 3D numerical simulations.
The global parameters include the momentum and kinetic energy of  shocked 
atomic/molecular gas and also the thermal energy of hot gas. 
These global parameters are mostly 
obtained from the literature but, for some SNRs, they are derived in this work. 
The empirical measurement is not straightforward because 
we cannot observe all the material engulfed by SN blast wave.
This is particularly serious for the \schi\ shells because  
we usually observe only the fastest expanding portion of 
the shell toward or away form us unambiguously. 
(The \schi\ emission line in the Galactic plane is 
very broad due to the Galactic rotation.)
Therefore, the estimates of 
momentum and kinetic energy of expanding \schi\ 
shells are obtained from an analysis where it is assumed that the 
\schi\ shell is thin and spherical and that the expansion 
velocity of the shell material is either constant or varying radially. 
The systematic uncertainties 
arising from this `{\em thin-shell analysis}' might 
depend on the complexity of the environment, and it is difficult to estimate the uncertainties 
for individual SNRs. 
In principle, we can apply the same thin-shell analysis to mock observations from 
simulations of SNRs, and, by comparing the result with that from observations
we can infer the properties of multiphase gas in the environments of individual SNRs 
and/or we can validate theoretical SNR models.
An effort of this kind would, however, require extensive 
(and numerically costly) parameter space investigation  
with simulations, given the large possible range of 
multiphase environments  of SNe.
As a first step, in this work we take a simpler approach and 
compare the global parameters derived from the thin-shell analysis 
to the {\em total} momentum and {\em total} kinetic/thermal 
energy of radiative SNRs obtained from numerical simulations.
We discuss the uncertainties in the comparison 
arising from applying the thin-shell analysis 
for a mock \schi\ 
observation of an SNR simulated in the two phase 
ISM (KO15) as a case study.

The organization of the paper is as follows. 
In \S~\ref{sec:obs}, we first survey the available observations of radiative SNRs and 
summarize the parameters of their expanding shells. 
We show that there are seven SNRs (including two old SNR candidates) 
with \schi\ shells expanding at $\ge 50$~\kms.  Four of them are known to be  
interacting with molecular clouds, and we also summarize the momentum and kinetic energy 
of shocked molecular gas in these SNRs as well as 
the thermal energies of the seven SNRs obtained from X-ray observations. 
In \S~\ref{sec:snr_prop}, we compare the observed properties of 
radiative SNRs with the results from numerical simulations and analyze them.
In \S~\ref{sec:discussion}, we discuss the results 
and caveats, including the uncertainty in the comparison arising 
from applying the thin-shell analysis.
And in \S~\ref{sec:conclusion}, we conclude our paper.

\section{Observed Properties of Radiative SNRs with \schi\ Shells}\label{sec:obs}

In this section, we present basic observed properties of seven radiative SNRs
that possess HV \schi\ shells, mainly from the literature.  We then derive momentum and
kinetic energy associated with the \schi\ shells and the molecular gas interacting
with the SNRs. We also derive thermal energy of the hot gas from X-ray
observations. Although we are not reanalyzing all the observations in this paper (we do for some of them; see Appendix), we summarize the procedures taken to obtain basic quantities
to clarify where the uncertainties arise from. 

\subsection{\schi\ Shells and Their Momentum and Kinetic Energy}\label{sec:HI}

There have been systematic searches for radiative \schi\ shells
associated with the {\em known} Galactic SNRs in the northern and southern skies \citep{koo91, koo04b}. 
These surveys detected HV \schi\ gas in 25 SNRs, but because of 
low angular resolution (i.e., FWHM$=16'$--36$'$), their association needed confirmation
except for the large SNRs, e.g., HB 21 of size $120'\times 90'$.
Among the 25 SNRs, the SNRs confirmed by 
high-resolution observations are  
CTB~80 \citep{koo90}, W44 \citep{koo95a}, W51C \citep{koo97a},
and IC~443 \citep{giovanelli79, braun86, lee08}.
\cite{park13} carried out a systematic search for fast-expanding \schi\ shells
toward the SNRs in the Galactic longitude range $\ell\approx$ 32\arcdeg\ to 
77\arcdeg\ using the data from the Inner-Galaxy Arecibo L-band Feed 
Array (I-GALFA) \schi\ survey. Their high-resolution study confirmed the \schi\ shell in   
G54.4$-$0.3 but did not detect additional SNRs with radiative shells.
So there are six SNRs (including HB 21) 
with their expanding shells studied with high enough spatial resolution. 
The expansion velocities of expanding \schi\ shells range 59--135 \kms. 
Among these, the \schi\ shell in W51C is a partial arc along a large molecular cloud 
well inside the X-ray/radio SNR that could be the result of multiple SNe (\citealt{koo95,koo97a}; cf. \citealt{tian13}),
so the morphology of the SNR-\schi\ shell system does not fit into a radiative SNR 
and we exclude the SNR W51C in our analysis.

Another approach is to search for 
localized HV \schi\ features in {\em unbiased} \schi\ surveys of the Galactic plane 
that are not directed specifically at known SNR positions
\citep{koo04a,kang07,kang14}.  Those HV features could be  
radiative shells associated with old, radio-faint SNRs not identified in radio continuum. 
\cite{koo04a} noted that there are many 
small, faint, HV “wing-like” features extending to velocities well beyond
the maximum or minimum permitted by Galactic rotation
in large-scale $(\ell, v)$ diagrams of \schi\ 21 cm line emission in
the Galactic plane and proposed that 
some of these ``forbidden-velocity wings (FVWs)'' may represent the
expanding shells of “missing” SNRs that are not included in existing SNR catalogs.
\cite{koo06} have carried out high-resolution
\schi\ line observations toward one of FVWs and showed that 
it is indeed a rapidly expanding (80 \kms) \schi\ shell 
and that its physical parameters are consistent with an old SNR (G190.2+1.1). 
\cite{kang12} studied another FVW around \schii\ regions 
at $\ell\sim 173\degr$ and concluded that it is probably an expanding shell associated with 
an old SNR (G172.8+1.5). These two fast-expanding \schi\ shell objects are   
old SNR {\em candidates} because their nature as an SNR has not been confirmed.
In what follows, however, we will  include these two candidates in the analysis.  

In summary, there are seven SNRs (including two 
old SNR candidates) with radiative shells expanding at $\ge 50$ \kms~(Table~\ref{tbl-1}).  
These middle-aged ($t\sim 10^4$--$10^5$ yrs) remnants are preferred for measuring 
the momentum injection quantitatively, since they are old enough to have reached terminal 
momentum, but not so old that they have started to merge into the background ISM.
There have been studies reporting the detection of almost stationary \schi\ shells or \schi\ bubbles
associated with SNRs (e.g., \citealt{kothes05}; \citealt{cazzolato05}; 
see also references in \citealt{koo04b}).
But it is difficult to obtain reliable shell parameters for such 
low-velocity \schi\ shells, and we limit our study to the seven SNRs in Table~\ref{tbl-1}.
The parameters in Table~\ref{tbl-1} are from the literature except 
for HB 21, which are derived in this work (see Appendix A). 

In Table~\ref{tbl-1}, radius ($\rshell$), expansion velocity ($\vshell$), and 
\schi\ mass ($\mshell$(\schi)) are obtained directly from radio continuum, \schi\ 21 cm line, and infrared  
observations. The method varies for SNRs, which are described briefly in the following. 
More details may be found in the cited references as well as in Appendix A.  

\begin{itemize}

\item[]{(1)} W44, G54.4$-$0.3, HB21: Middle-aged SNRs with possibly a complete \schi\ shell. 
These SNRs appear circular or elliptical loops in radio continuum 
(e.g., see Figure~\ref{fig:w44-ic443}), and 
the receding portion of an expanding \schi{} shell is 
clearly visible inside the remnant including the cap at the center (\citealt{koo91,park13}; see also Appendix A).  
In W44, both the receding and approaching portions are detected. 
For these SNRs, the radius of the \schi\ shell is assumed to be the same as the radio continuum loop, while   
the expansion velocity and the mass of the shell are 
derived from the average \schi{} spectrum of the SNR by assuming a spherical 
thin shell of radius $\rshell$ where $\rshell$ 
is the geometrical mean radius for W44 and HB21 which are elliptical.

\item[]{(2)} IC 443: Middle-aged SNR with a partial \schi\ shell along the SNR boundary. 
The SNR appears as two circular loops of different radii in radio continuum, and 
the portions of an expanding \schi\ shell are visible along the boundary of the smaller 
and brighter circular loop in the east (\citealt{lee08}; see also Figure\ref{fig:w44-ic443}). 
In the northern area, both approaching and receding portions of the shell are detected, while 
in the southern area, where the SNR is interacting with a MC, 
most of the mass resides in the approaching portion. 
$\mshell$(\schi) is derived from the mass distribution 
of shocked \schi\ gas $M(v_\ell)$ where $v_\ell$ is the line of sight (LOS) velocity, 
either by assuming a Gaussian profile (northern area) or 
by using the emission profile of shocked molecular gas as a template (southern area) \citep{lee08}.
The expansion velocity $\vshell$ of the shell is assumed to 
be equal to the shock velocity derived from modeling the optical filaments \citep{alarie19}, while    
the radius of the \schi\ shell is assumed to be the same as that of the smaller 
radio continuum loop.

\item[]{(3)} CTB 80 and G190.2+1.1: Old SNRs with possibly complete shells.
The SNR appears either irregular (CTB 80) or is 
not visible (G190.2+1.1) in radio continuum, but 
the receding portion of an expanding \schi{} shell is 
clearly visible inside the remnant including the cap at the center. 
In CTB 80, a partially complete, circular infrared loop of shock-heated dust is present, 
and the radius of the \schi\ shell is assumed to be the  same as this infrared shell \citep{koo90}.
The expansion velocity of the shell is derived from the \schi\ line profile at the center \citep{koo90,park13}.
In G190.2+1.1, which is an old SNR candidate of elliptical shape, 
the projected size 
of the shell varies with 
LOS velocity, which was used to derive the 
geometrical mean radius and expansion velocity of the shell \citep{koo16}. 
The mass of the shell in these SNRs 
is derived from the average \schi{} spectrum of the SNR by 
modeling $M(v_\ell)$ with a Gaussian profile 
assuming spherical symmetry \citep{park13,koo16}.

\item[]{(4)} G172.8+1.5: Old SNR with possibly a complete shell.
This is an old SNR candidate that appears as a faint 
loop of horseshoe-shape  in radio continuum \citep{kang12}.
Diffuse and clumpy \schi\ emission features with a velocity structure   
indicating the receding portion of expanding \schi{} shell are present. 
The expansion velocity and the mass of the shell are  
derived from the average \schi{} spectrum of the SNR by assuming a 
thick spherical shell of geometrical mean 
radius $\rshell$ with a velocity gradient inside \citep{kang12}.
The radius of the \schi\ shell is assumed to be the same as the radio continuum loop.

\end{itemize}

The momentum ($\pshell$) and 
the kinetic energy ($\ekshell$) of the shell are derived from $\mshell$(\schi) and $\vshell$;  
$\pshell=\mshell\vshell$ and $\ekshell=\mshell\vshell^2/2$ 
where $\mshell(=1.4\mshell$(\schi)) is the mass of shell including 
the contribution from He assuming cosmic abundance. 
The parameter $\tshell$ is the characteristic age  of the shell defined as 
\begin{equation}\label{eq:tsh}
\tshell \equiv 0.3 {\rshell \over \vshell}.
\end{equation}
This characteristic age $\tshell$ should be close to the age of the SNR if 
the SNR is in the pressure-driven snowplow (PDS) phase. 
But the expansion parameter $\eta$ ($R\propto t^\eta$) that appears as a prefactor in Equation (\ref{eq:tsh})
could vary from 0.4 to 0.25 depending on the evolutionary state of the SNR 
so that it could be slightly different from the true age.
According to Table~\ref{tbl-1}, the radiative \schi\ shells of our sample of SNRs 
have $\pshell=(0.5$--$4.5)\times 10^5$ \Msun~\kms, 
$\ekshell=(0.4$--$3.5)\times 10^{50}$ erg, and 
$\tshell=(0.3$--$3.4)\times 10^5$~yr. 

There are some caveats in using the physical parameters in Table~\ref{tbl-1}. 
First, the {\em observed} \schi{} mass is only a small fraction of the total shell mass 
because only the portions of the shell with high enough LOS velocities are visible. 
So the parameters $\mshell,\pshell,\ekshell$ are obtained by applying 
the thin-shell analysis where it is assumed that the \schi\ shell is 
thin and spherical and that the expansion velocity of the shell material is 
either constant or varying radially
depending on the observed properties of the \schi\ 21 cm emission.
The systematic uncertainties arising from 
this thin-shell analysis might depend on the complexity of the environment, 
and it is difficult to quantify the errors for individual SNRs. 
(Note that the errors in Table \ref{tbl-1} are formal errors 
associated with the measurement errors.) 
The parameters $\mshell,\pshell,\ekshell$ in Table~\ref{tbl-1} should be 
understood as those derived from the thin-shell analysis.  
We will discuss the systematic uncertainties in the thin-shell analysis 
in \S~\ref{sec:diss_error}.
Second, the distance ($d$) to the SNRs is uncertain.
The uncertainty in the distances in Table~1 is probably $\simlt 30$~\% 
except for G 190.2$+$1.1. 
For G190.2$-$1.1, there is only circumstantial evidence and 8 kpc has been suggested to yield  
the SN explosion energy of $10^{51}$~erg \citep{koo06}.  
Note that $\mshell\propto d^2$, so that   
the uncertainty of 30~\% in $d$ yields $\pshell$ and $\ekshell$ 
uncertain of 60\%.

\subsection{Shocked Molecular Gas and Its Momentum and Kinetic Energy}\label{sec:mol}

Among the seven SNRs in Table~\ref{tbl-1}, four SNRs are known to be interacting with MCs: W44, G54.4$-$0.3, HB 21, and IC 443 (see Note in Table~\ref{tbl-2}). 
W44 and IC 443 are the two prototypical SNRs interacting with large MCs (Figure~\ref{fig:w44-ic443}). 
The presence of both shocked atomic (\schi) and shocked molecular (e.g., CO and HCO$^+$) gases 
in these SNRs can be explained by SN explosion inside a clumpy or turbulent MC. 
In such case, the shocked molecular gas usually represents 
non-dissociative shock propagating into the dense clumps 
while the HV \schi\ gas is from a 
radiative atomic shock propagating into the rather diffuse interclump gas
\citep{chevalier99,reach05,koo14,zhang19}.
This model can also explain the centrally-brightened X-ray emission (see \S~\ref{sec:hot}).
In addition to the clumpiness, the molecular line emission confined to 
one side of the SNR suggests a surrounding medium with a  
large-scale density gradient where thermal conduction can substantially alter 
the interior structure of an SNR \citep{cox99}.
In \S~\ref{sec:discussion}, we will discuss the effect of such complex environment in 
analyzing the observed global parameters. 
In the other SNRs, the shocked molecular gas is either not prominent or has 
not been detected (see Note in Table~\ref{tbl-2}). 

The momentum and kinetic energy of shocked molecular gas can be derived from 
the molecular line emission from shocked molecular gas. 
Table~\ref{tbl-2} summarizes the results available in the literature. 
In Table~\ref{tbl-2}, the mass of H$_2$ `shell' ($\mshell$(H$_2$)) represents the mass of shocked molecular 
H$_2$ gas obtained from the emission line flux of molecules such as HCO$^+$ and CO, and  
there is a large ($\sim 50$~\%) uncertainty in scaling the emission line flux to H$_2$ mass. 
The expansion velocity ($\vshell$) is derived from a fit to the velocity structure (W44, IC 443) 
or from the mean velocity of shocked molecular clumps (HB 21).
The momentum ($\pshell$) and the kinetic energy ($\ekshell$) of the shocked molecular gas are 
derived from $\mshell$(H$_2$) and $\vshell$ as in \schi\ shell, i.e.,   
$\pshell=\mshell\vshell$ and $\ekshell=\mshell\vshell^2/2$ where $\mshell=1.4\mshell$(H$_2$). 
According to Table~\ref{tbl-2}, 
in W44 and IC 443, the momentum of shocked molecular gas is comparable to or even larger than  
that of the \schi\ shell (1.8 and $0.7\times 10^5$ \Msun~\kms), while its 
kinetic energy is relatively small ($\simlt 0.2\times 10^{50}$ erg).  
For G54.4$-$0.3, large MCs are around the SNR and it has been proposed that 
this `molecular shell' is expanding at 5~\kms, in which case 
its momentum would be $\sim 1\times 10^6$~\Msun~\kms\ \citep{junkes92a}. 
But there is no convincing evidence that the large MCs are shocked and/or expanding. 
Instead, there are small, thin \h212\ emission filaments 
implying a small amount of shocked molecular gas (see Note in Table~\ref{tbl-2}).
For HB 21, the momentum and kinetic energy of shocked molecular gas are not significant. 
For the remaining three SNRs, the shocked molecular gas has not been detected, 
so that the momentum and kinetic energy of shocked molecular gas might not be significant. 

\subsection{Hot Gas and Its Thermal Energy}\label{sec:hot}

Six SNRs are detected in X-ray, and the physical parameters of 
the X-ray emitting hot gas including its thermal energy are listed in Table~\ref{tbl-3}. 
All results are obtained from the literature except  
G54.4$-$0.3 and CTB~80, for which we derive it in this paper 
(see Appendices \ref{sec:appendixB1} and \ref{sec:appendixC}).
Table~\ref{tbl-3} shows that the mean density of the hot gas in six SNRs 
is in the range of 0.006--2~cm$^{-3}$, while the 
temperature of hot gas is $< 1$ keV. 
The thermal energy of the hot gas is in the range of 
0.6--$8 \times 10^{50}$~erg.
      
The parameters in Table~\ref{tbl-3} are derived from spectral analysis  
except G172.8$+1.5$ where the X-ray emission is seen only in the broad band {\em ROSAT} 
all-sky X-ray map. The X-ray spectral analysis 
provides the electron temperature ($T$) and the normalization factor proportional to 
$EM/4\pi d^2$ where $EM=\int n_e n_{\rm H} dV$ ($n_e$, $n_{\rm H}$ 
= electron and hydrogen densities) is 
the volume emission measure and $d$ is the distance to the source. 
Therefore, the average electron density in terms of $EM$ is given by     
$\bar n_e = g_e^{1/2} (EM/V_{EM})^{1/2}  \propto d^{-1/2}$ 
where $g_e = n_e/n_{\rm H}\approx 1.2$ and  
$V_{EM}$ is the volume of the X-ray emitting hot plasma.
The uncertainty of 30~\% in $d$ yields $E_{\rm th}$ uncertain by a factor of $\sim 2$.
It is worthwhile to note that $V_{EM}$ is often considerably smaller 
than the SNR because the area used for the X-ray spectral analysis is 
limited. For G54.4$-$0.3 and CTB 80, for example, the area is limited to the 
X-ray bright part of the SNRs and we, for a simple approximation, 
assumed that the hot plasma is uniformly filling a cylindrical volume with a path length (along the line of
sight) corresponding to the physical diameter of the spherical SNR
(Appendices~\ref{sec:appendixB1} and \ref{sec:appendixC}).
For the other sources, see Note in Table~\ref{tbl-3}.
A major uncertainty in the {\em total} thermal energy of an SNR 
arises from the geometry of hot gas. 
The SNRs in general have non-uniform X-ray 
brightness, which might be due to the combination of 
temperature/density/metallicity structure and varying absorbing columns. 
In W44, for example, it has been shown that 
the X-ray emission filling the interior is faint near the SNR shell 
because of low electron temperature and large absorbing columns 
(see Fig.~\ref{fig:w44-ic443}; \citealt{okon20}). 
For this study, we derive the thermal energy of SNRs 
assuming that the entire interior volume of 
the SNRs is filled with hot gas in pressure equilibrium, i.e.,  
$E_{\rm th} \approx 3 {\bar n_e} k_B T V_s$ where 
$k_B$ is the Boltzmann constant and 
$V_s (\equiv 4\pi R_s^3/3)$ is the total volume of the SNR. 
(For IC 443, we assumed two hemispherical spheres following 
\citealt{troja06}. Some notes on individual SNRs are given in Table~\ref{tbl-3}.)  
We note that our assumption of  a spherical volume of radius $R_s$ 
might be an over-simplification to yield an overestimate of the volume in general, 
although the pressure equilibrium might be a reasonable assumption.
The systematic uncertainties in the   
thermal energy estimate due to the uncertainties in the geometry of an SNR, 
e.g., the extent along the LOS and the volume filling factor of hot gas, 
are difficult to estimate. 
The thermal energies in Table~\ref{tbl-3} 
should be understood as those obtained by assuming 
a spherical volume filled with hot gas in pressure equilibrium. 
If the volume filling factor of hot gas is $\sim 0.5$, 
then the thermal energy would be $\sim 1/2$ of that in the table
(see \S~\ref{sec:diss_env} and \S~\ref{sec:diss_error}).

It is worth making a comment on the X-ray morphology of SNRs.
As we can see in Figure~\ref{fig:w44-ic443}, 
the X-ray brightness in some SNRs is centrally peaked, not limb brightened 
as we expect in an `ideal' SNR.  
Essentially all six SNRs visible in X-ray have such X-ray morphology including 
CTB 80 with peculiar radio morphology and the old SNR G172.8+1.5.
They belong to the so-called ``thermal composite" or 
``mixed-morphology" SNRs (hereafter MM SNRs) which are the SNRs 
that are shell or composite type in radio but with 
centrally-brightened thermal X-rays inside the SNR \citep{rho98,lazendic06,vink12}.
Most of these MM SNRs are thought to be middle-aged SNRs interacting with dense ISM, 
e.g., molecular clouds \citep{zhang15,slane15,koo16}. 
W44 and  IC 443 are prototypical MM SNRs. 
The origin of the centrally-brightened 
thermal X-ray is not clear \citep{slane02,lazendic06,vink12,slane15}. The proposed scenarios are 
(1) absorption of soft X-rays from near the edge of the remnant 
by the large absorbing column density \citep{harrus97,velazquez04},
(2) large-scale conduction between the hot interior and the cool SNR shell 
\citep{cox99, shelton99,velazquez04,tilley06,okon20},
and (3) evaporation of small, dense clumps engulfed by SN blast wave 
\citep{white91,slavin17,zhanggao19}. 
In some SNRs, it has been found that heavy elements are enriched, 
which can partly responsible for the centrally-brightened X-ray emission
(see \citealt{lazendic06, vink12} and references therein).
In several MM SNRs, it has been also found that 
the hot plasma is overionized implying a rapid cooling,  
but its relation to the X-ray morphology is not clear (e.g., 
\citealt{slane15} and references therein).
Recently, there have been 2D and 3D numerical simulations in  
clumpy or turbulent media to understand the origin of  
MM SNRs \citep{slavin17,zhang19, zhanggao19}, 
but the role of thermal conduction in 
thermal energy evolution remains to be explored (see section~\ref{sec:diss_physics}).

\section{Integrated Properties of Radiative SNRs}\label{sec:snr_prop}

\subsection{Observed Global Parameters}\label{sec:obs_prop}

Table~\ref{tbl-4} summarizes the observed global parameters of radiative SNRs; $\pshell$, $\ekshell$, and $\eth$. 
For SNRs with shocked molecular gas, $\pshell$ and $\ekshell$ of the atomic (\schi), 
molecular (H$_2$), and atomic+molecular components are given.
The table shows that the total (\schi+H$_{2}$) momentum and kinetic energy of the radiative SNRs are in the range of 
$\pshell=$(1.1--4.5)$\times 10^5$ \Msun \kms\ and  
$\ekshell=$(0.6--3.5)$\times 10^{50}$ erg, respectively, while $\eth=(0.6-8)\times 10^{50}$~erg.   
In the table, $\nbarc$ is the density of hydrogen nuclei if the mass of shocked atomic/molecular gas were uniformly filling the 
spherical volume of radius $\rshell$.

If the ambient medium is isotropic, 
$\nbarc$ in the `\schi+${\rm H}_2$' column of Table~\ref{tbl-4} 
will be close to the mean density of the ambient medium,
in which SN blast waves propagate,
even if the medium is inhomogeneous.
It is, however, not unusual that the ambient medium has a large-scale non-uniformity for SNRs, e.g.,
SNRs produced inside or outside large MCs near their surface \citep[][and references therein]{tenorio-tagle85,jeong13,cho15,iffrig15}. 
For example, IC 443 in Figure~\ref{fig:w44-ic443} is interacting with a dense MC in the southern area 
and the shocked atomic and molecular gases are mainly confined to that area.  
The density in the northern area where the X-ray emission 
is prominent could be significantly lower than the total mean density. 
W44, on the other hand, appears to be located  
inside an MC \citep{chevalier99,zhang19}. But again, most of the cool, shocked  
molecular gas is confined to the eastern area \citep[e.g.,][see also Figure~\ref{fig:w44-ic443}]{sashida13}.
The mean ambient density inferred from \schi\ and X-ray observations 
($\simlt 10$~cm$^{-3}$, see \citealt{koo95,chevalier99,zhang19} and references therein) is indeed much lower than the average density
when the shocked atomic and molecular gases were uniformly filling the 
spherical volume of the SNR (i.e., 50--70 cm$^{-3}$, Table \ref{tbl-4}).
One therefore needs to be cautious in using the characteristic densities in Table~\ref{tbl-4}, 
particularly for SNRs interacting with MCs (see Note in Table~\ref{tbl-2} 
for a brief description of individual SNRs).

Bearing the above caveats in mind, we show in Figure~\ref{fig:obs} the observed
momentum and energy of the SNR as a function of $\tshell$. For comparison, we
also plot theoretical evolutionary tracks 
obtained from 1-dimensional hydrodynamic simulations using Athena++ 
\citep{stone20}\footnote{https://princetonuniversity.github.io/athena/}, in
which we follow the evolution of SNRs in a uniform medium with
spherical symmetry including radiative cooling and thermal conduction
\citep[see][for details]{el-badry19}.
For theoretical curves, we use the true age of the SNR $\tshell=t$ not
$\tshell=0.3\rshell/\vshell$ as the $x$-axis, e.g.,  we plot
$\pshell(\tshell=t)$ not $\pshell(0.3\rshell/\vshell)$. 
The difference in momentum by using  
$t$ not $\tshell$ is $\simlt 10$\% 
before the shell formation time and is negligible after that. 
For kinetic energy, the difference is almost negligible 
before the shell formation time and slowly increases to 5 \% until the final stage. 
Considering the uncertainty in observed parameters and the idealization of the
spherically symmetric simulations, we simply use $t$ for the theoretical
curves.

In Figure~\ref{fig:obs}, it is remarkable to see that the observed momenta of 
seven SNRs are all very close to the theoretical result for the canonical SN explosion 
energy $E_{51}\equiv(\esn/10^{51}~{\rm erg})=1.0$.
W44, for example, has a momentum corresponding to an SNR  
in the momentum-conserving phase in  
dense environment of mean ambient density $n_0 \equiv(\nbg/1~{\rm cm}^{-3}) \simgt 10$ 
where $\nbg$ is the hydrogen number density of the uniform background medium. 
As we pointed out above, 
the ambient density may vary considerably for a given SNR as well as from SNR to SNR. 
But, since the SNR momentum is insensitive to ambient density but  
sensitive to SN explosion energy, i.e., $\pshell\propto n_0^{-1/7}E_{51}^{13/14}$ \citep[e.g.,][]{cioffi88,thornton98,kim15}, 
the SN explosion energy smaller or greater than the canonical value by a factor of 
2--3 can explain the observed range of the SNR momentum.   
The observed range of the SNR kinetic energy 
(0.6--3.5$\times 10^{50}$ erg) also matches the theoretical result 
for $E_{51}\sim 1$. 
IC 443 appears to be far from the  theoretical curve with $E_{51}=1.0$ but, 
since its mean ambient density is $\sim 60$~cm$^{-3}$, $E_{51}\sim 0.5$ can 
explain the kinetic energy.

The thermal energy has a relatively large scatter;  
CTB 80 has thermal energy of 6$\times10^{49}$~erg while 
the other SNRs have thermal energies much larger than this (2--8$\times10^{50}$~erg). 
The large scatter might be partly because the thermal energy sensitively depends 
on both $n_0$ and $\esn$, i.e., $\eth\propto E_{51}^{1.35} n_0^{-0.86}$ (\S~\ref{sec:comp_sims}). 
Therefore, again the SN explosion energy smaller or greater than the canonical value by a factor of 
2--3 can explain the observed range of the SNR thermal energy, although 
the corresponding density does not yield the observed momentum and kinetic energy. 
One thing to notice is that the observed thermal energies are generally greater than the theoretical 
result for $E_{51}=1$ in contrast to the observed shell momenta and kinetic energies 
which are generally smaller than the theoretical result for $E_{51}=1$.
HB 21, for example, 
has thermal energy that is not particularly large (i.e., $\eth \sim 2\times 10^{50}$~erg) but, since 
the SNR thermal energy decreases  slightly super-linearly
with time after the shell formation, 
it is still significantly greater than the expected thermal energy 
for $E_{51}=1$ and $n_0=1$. 
The same is true for W44 and IC 443; their thermal (kinetic) energies are 
larger (smaller) than those expected for $E_{51}=1$ and $n_0=10$.
The only SNR that has thermal energy smaller than the theoretical result 
for $E_{51}=1$ is CTB 80.
We consider that this discrepancy is due to the complex environments of the SNRs, and 
we will discuss this in \S~\ref{sec:diss_env} and \S~\ref{sec:diss_error}.

\subsection{Comparison with Simulations}\label{sec:comp_sims}

The real ISM is highly inhomogeneous and structured. Therefore, the comparison
with the model SNR evolution within a uniform medium is possibly too
simplistic. In this subsection, we present evolutionary tracks of the
integrated SNR properties from a variety of numerical simulations to understand
systematic uncertainties in theoretical models.

To  explore the  role of ambient conditions, we use additional 3D hydrodynamic simulations including two
kinds of initial background medium states: (1) inhomogeneous background medium as a
result of thermal instability (TI model; see KO15) and (2) contacting two
homogeneous slabs (TS model; see \citealt{cho15}). Figure~\ref{fig:snapshot}
shows example snapshots from TI (top) and TS (bottom) models. For the
TI model, we first run a simulation with initially thermally unstable
equilibrium state and let it evolve toward a nonlinear saturated state of
thermal instability \citep[][]{field65,piontek04,kim08,choi12,inoue15}. The
resulting medium consists of two distinct thermal phases in the pressure
equilibrium, namely cold neutral medium and warm neutral medium (CNM and WNM,
respectively), with density and temperature contrasts about 100. The WNM is
volume filling ($\sim90$\%), while the CNM comprises most of the mass ($\sim80$\%).
As a result, the TI model provides a natural way to set up a clumpy ISM,
with characteristics determined by cooling and heating processes of the ISM
\citep[e.g.,][]{field65,field69,wolfire95,bialy19}. We use 10 realizations of
the TI model with mean hydrogen density of $\bar\nbg=1$~cm$^{-3}$ and 
$\bar\nbg=10$~cm$^{-3}$,
while we show an example for the $\bar\nbg=1$~cm$^{-3}$ case in the top row of
Figure~\ref{fig:snapshot}. We refer the readers KO15 for complete descriptions
of model setups and more comprehensive parameter study.  The TS model is
characterized by the densities of two media (which are fixed to $\nbg=1\pcc$
and $100\pcc$ here) and the explosion depth from the contacting surface toward
the denser medium 
$h$.  We present seven different explosion depths varying
from $h=-1$, 0, 1, 2, 2.5, 3, and 4 pc, while we show an example for the
$h=2{\rm\, pc}$ case in the bottom row of Figure~\ref{fig:snapshot}.  We refer
the readers \cite{cho15} for complete descriptions of model setups and more
comprehensive parameter study.\footnote{Note that the results presented here
are resimulated using the code and method described in KO15.}

Note that the background medium realizations presented here are by no means a
complete census of SN explosion sites. We show two representative examples to
demonstrate possible variations of SNR evolution due to background medium
states, but there are many other possible realizations
\citep[e.g.,][]{li15,martizzi15,walch15,iffrig15,zhang19}.

Figure~\ref{fig:sims} presents evolutionary tracks of the integrated SNR
properties from all numerical simulations (left column for (a) total radial
momentum, (c) kinetic energy, and (e) thermal energy), including the 1D
spherically symmetric simulations in a homogeneous medium presented in
Figure~\ref{fig:obs}. In Figure~\ref{fig:sims}, the physical quantities are
normalized by the
parameters at the shell formation, i.e., 
\begin{eqnarray}
\tsf & = & 4.4\times 10^4{\,\rm yr}\,E_{51}^{0.22} n_0^{-0.55}, \label{eq:tsf} \\
\psf & = & 2.17\times 10^5{\, M_\odot\,{\rm  km~s}^{-1}}\,E_{51}^{0.93}n_0^{-0.13} 
\label{eq:psf}
\end{eqnarray} 
(Eqs. 7 and 17 in KO15) and the explosion energy used in the simulations. 
Note that $n_0=n_{\rm bg}/1~{\rm cm}^{-3}$ and
that we again use the true age ($t$)  of SNR for the $x$-axis. 
For $n_0$, we use the mean density of the whole SNR bubble
defined by gas with $T > 2 \times 10^4$~K or $|v| > 1$~\kms\ (e.g., KO15).
For the TI models, $n_0$ is nearly constant over time, 
while it varies with time for the TS models. 
For reference, $\tsf=156$, 44, 12, and 3.5 kyr for $n_0=0.1$, 1, 10,
and 100, respectively. 
Noticeably, simulations with different background realizations follow
very similar evolutionary tracks for momentum and kinetic energy, while the
thermal energy evolution can vary significantly.

The model prediction of total radial momentum and kinetic energy (after the
shell formation) for a given explosion energy is quite robust irrespective of
the complexity of the explosion environment. This is mainly due to (1) weak dependence 
and no dependence on the mean density respectively of momentum and kinetic energy at shell formation 
and (2) a minimal increase/decrease of momentum and kinetic
energy after the shell formation. Despite a wide variety of background
conditions considered here, total momentum/kinetic energy in simulations after
shell formation differs by less than a factor of 2. As a consequence, total
momentum and kinetic energy measured theoretically for radiative SNRs are not sensitive to
the mean density but to the explosion energy.

This motivates us to map the global properties of observed SNRs into
the normalized quantity plane in the right column of Figure~\ref{fig:sims}.
To calculate the normalization factors $\tsf$ and $\psf$ from Equations (\ref{eq:tsf}) and (\ref{eq:psf})
for observed SNRs, we adopt $\nbarc$ in Table~\ref{tbl-4} and the canonical explosion energy $\esn=10^{51}$~erg.
For SNRs with associated MCs, two observation points are shown not only for 
total radial momentum and kinetic energy excluding and including H$_2$ contributions
but also for different $\tsf$ and $\psf$ derived from $\nbarc($\schi$)$ and $\nbarc($\schi$+{\rm H}_2)$
as open and filled circles, respectively, connected by a dotted line.
As we adopt a fixed value for $\esn$ and the uncertainty in $\nbarc$ is dominated by systematic errors,
the uncertainties of observed SNRs in Figure~\ref{fig:sims} are not well defined and difficult to present.
We instead show two vectors in panels (b), (d), and (e) to present how the observation points
would move if $\nbarc$ or $\esn$ is increased by factor of two.
For momentum, the $\nbarc$ vector is almost parallel to the theoretical curve (dashed line; see below)
after shell formation time. 
For kinetic energy, the $\nbarc$ vector is horizontal, while 
the theoretical evolution track falls sub-linearly with time.
This allows us to derive the explosion energy of SNRs using momentum and
kinetic energy reliably \emph{assuming the momentum and kinetic energy inferred from the described observations comprise the majority of these quantities.} 

To get the explosion energy, we use analytic expressions for radius
and momentum evolution during the Sedov-Taylor and post shell formation stages
(e.g., KO15): 
\begin{eqnarray}
\rshell &= &
\left\{
\begin{array}{lc}
    \rsf (t/\tsf)^{2/5} & \quad\textrm{if $t<\tsf$} \\
    \rsf (t/\tsf)^{2/7} & \quad\textrm{otherwise,}
\end{array}\right.\label{eq:rsh}\\
\pshell &=& 
\left\{
\begin{array}{lc}
    \psf (t/\tsf)^{3/5} & \quad\textrm{if $t<\tsf$} \\
    \psf (1+0.62(1- (t/\tsf)^{-6/7})) & \quad\textrm{otherwise,}
\end{array}\right.\label{eq:psh}
\end{eqnarray}
where 
\begin{equation}
\rsf = 22.6 E_{51}^{0.29} n_0^{-0.42} \quad{\rm pc} \label{eq:rsf}
\end{equation}
is the radius at the time of shell formation. Then, the kinetic energy can be
calculated $\ekshell = \pshell^2/(2\mshell)$, where
$\mshell=(4\pi/3)\rshell^3\rho_{\rm bg}$. 
As $t \gg \tsf$, the terminal momentum approaches to $1.6\psf$ with dependence on explosion energy and ambient density following
\begin{equation}\label{eq:psh_asym}
\pshell
  \approx 3.47\times 10^5 {\,M_\odot\,{\rm  km~s}^{-1}}\, E_{51}^{0.93}n_0^{-0.13}
 \qquad\textrm{if $t \gg \tsf$},
\end{equation}
and the kinetic energy drops nearly linearly with age as
\begin{equation}\label{eq:eksh_asym}
\ekshell
\approx 3.53\times 10^{50}{\,\rm erg}\, E_{51}^{1.18} n_0^{-0.47} 
\left(t\over 10^5~{\rm yr} \right)^{-0.86}  \qquad\textrm{if $t \gg \tsf$}.
\end{equation} 
Note that  these expressions are derived by assuming that the shell is driven
by the pressure of hot interior gas. 
In the TI models, it was shown that the hot gas pressure drops 
below the shell gas pressure later in 
the evolution, so that the final momentum of the shell
is slightly less than that in Equation (7) (see Eq. 29 of KO15).
We plot the theoretical curves in
Figure~\ref{fig:sims}, which are in excellent agreement with the 1D simulations
(blue and green lines) and generally consistent with the 3D simulations within
a factor of 0.6--1.2 as demonstrated by the gray shaded region in
Figure~\ref{fig:sims}. 
For the TI models, simulation gives overall smaller momentum at the 
same $t/\tsf$ mainly because the `real'
shell formation time (defined by the time at which the hot gas mass begins to decrease) for the TI models is about two times longer than that of 
the uniform medium (see Eq. 30 in KO15) 
since shell formation is delayed in the volume filling WNM. 
With $\tsf$ determined by the simulation for the TI models (shifting the curves to the left), the momentum evolution is more or less similar with the theoretical model. The kinetic energy is smaller in the TI models because the higher density, lower velocity gas (originally from the CNM), which contains the same momentum, has smaller kinetic energy. 
For the TS models, the SNR evolution is characterized not only by the properties at the shell formation time of the one medium but also at the time at which SNR breaks out to the low density medium. Figure 4, however, shows that the analytic expressions of $\pshell$ and 
$\ekshell$ (Eqs. \ref{eq:psh_asym} and \ref{eq:eksh_asym}) 
are also generally consistent with the result of the TS models. 

In Table~\ref{tbl-5}, we present the SN explosion energies $\esnpsh$ and $\esnksh$ derived by matching the
observed momentum  and kinetic energy 
to those of Equations (\ref{eq:psh_asym}) and (\ref{eq:eksh_asym})  
respectively 
for given mean ambient density and age presented in Table~\ref{tbl-4}. 
Although the mean ambient density is very difficult to
estimate from observations, any uncertainty in the ambient density only marginally
affects the derived explosion energy of SNe as long as observed momentum and
kinetic energy are the majority. 
As demonstrated by the gray shaded area in Figure~\ref{fig:sims}, 
there are inherent uncertainties in estimating $E_{\rm SN}$
using a theoretical model related to the complexity of the explosion site.
Figure~\ref{fig:sims}(b) and (d) 
show that all SNRs except CTB 80 are broadly
consistent with the canonical explosion energy of $E_{51}=1$ unless the
observed measurements of momentum and kinetic energy are significantly
underestimated. Both $\esnpsh$ and $\esnksh$ are
within $0.3-1.4\times10^{51}\erg$.

In contrast to momentum and kinetic energy, thermal energy evolution of
radiative SNRs is sensitive to both the mean density (or evolutionary stage)
and explosion energy since the thermal energy is quickly radiated away after
the shell formation. If the background medium is inhomogeneous, the shocked
medium begins to cool earlier as the blast wave sweeps up denser medium. The
evolution of thermal energy is thus sensitive to the fractions of swept-up gas
at different densities (or background medium structure). If the density
structure is more or less isotropic (as in the TI model; see top row of
Figure~\ref{fig:snapshot}), the CNM causes cooling while the WNM is still in
the Sedov-Taylor stage, but the overall cooling is actually delayed compared to
the uniform medium case with the same mean density since the WNM covers larger
volume so that the shell formation time is effectively longer.
If we substitute $\tsf$ determined by the simulation (Eq. 30 in KO15), 
the overall thermal energy evolution in the TI models is shifted to the 
left in Figure~\ref{fig:sims}(e). 
However, if the background medium is highly
anisotropic (as in the TS model; see bottom row of Figure~\ref{fig:snapshot}),
the evolutionary tracks of thermal energy cannot be described by a simple thin
shell evolution model with a single mean density. The TS model evolution shows
a two-step evolution characterized by two distinct volume filling densities at
different evolutionary stages. Thus, their evolutionary tracks can fill in a
large area in Figure~\ref{fig:sims}(e).

We repeat the similar exercise with the thermal energy to derive the explosion
energy of the observed SNRs (see Figure~\ref{fig:sims}(f)). We use the
theoretical model for the thermal energy evolution after the shell formation
presented in KO15 (see their Eq. 26):
\begin{equation}\label{eq:eth}
\eth =
\left\{
\begin{array}{lc}
    7.17\times10^{50}\erg\,E_{51} & \quad\textrm{if $t<\tsf$} \\
    5.74\times10^{50}\erg\,E_{51} (\rshell/\rsf)^{-2}(t/\tsf)^{-1} & \quad\textrm{otherwise}.
\end{array}
\right.
\end{equation}
If we substitute $\tsf$ and $\rshell$ in Equations (\ref{eq:tsf}) and (\ref{eq:rsh}), thermal energy at $t>\tsf$ can be written as 
\begin{equation}\label{eq:eth_asym}
\eth\approx 1.58\times 10^{50}\erg\, E_{51}^{1.35} n_0^{-0.86} 
\left(t\over 10^5~{\rm yr} \right)^{-1.57}
\quad\textrm{if $t>\tsf$}.
\end{equation}
Note that the simulation evolutionary tracks from the TI and TS models in Figure~\ref{fig:sims}(e) generally are declining more rapidly than the dashed line representing Equation (\ref{eq:eth}). It is expected to be caused by the enhanced cooling at the hot-cold interface due to both physically increased interface surface area in the inhomogeneous medium and numerically broadened interface \citep[see][for related discussions]{gentry19,el-badry19}. The gray shaded region in Figure~\ref{fig:sims}(e) now covers 0.1-3$\times$ the theoretical model to enclose the large range of evolutionary tracks from different background medium realizations. The theoretical model uncertainty for thermal energy is much larger than that of momentum and kinetic energy.

The SN explosion energy $\esnth$ derived by matching the
observed thermal energy to $\eth$ of Equations (\ref{eq:eth_asym}) 
is listed in Table~\ref{tbl-5}. Overall, the observed thermal energy requires
much larger explosion energy than that inferred from the momentum and kinetic
energy; $\esnth$ is larger than $\esnksh$ by a factor of $\ge 3$ for all
SNRs except CTB 80. 
This can be also seen in Figure \ref{fig:sims}(f), where 
four out of six SNRs are above the gray shaded area.
We will discuss this discrepancy in \S~\ref{sec:diss_env}.

\section{Discussion}\label{sec:discussion}

\subsection{Global Parameters and Environmental Effects}\label{sec:diss_env}

The comparison in Figures~\ref{fig:obs} and \ref{fig:sims} showed that 
the observation-based global parameters of SNRs (i.e., $\pshell$, $\ekshell$, and $\eth$)
are generally consistent with 
the 1D hydrodynamic simulations but at the same time that   
for most SNRs they cannot be explained 
by a single set of $\esn$ and $\nbg$.
One noticeable feature is that, 
for a given $\nbg$, the SN explosion energy derived from 
thermal energy $\esnth$ is generally larger than those derived from  
momentum and kinetic energy of the shell, $\esnpsh$ and $\esnksh$, while 
the latter two are almost comparable (Table~\ref{tbl-5}; see also Figure~\ref{fig:snenergy}).  
Five out of six SNRs with both $\esnth$ and $\esnksh$ estimated 
have $\esnth/\esnksh \ge 3 $. In particular, for the two 
prototypical SNRs interacting with large MCs, 
i.e., W44 and  IC 443, $\esnth/\esnksh$=6--8 
when we adopt $\nbarc$ corresponding to the density 
of the shocked \schi+H$_2$ gases averaged over the SNR volume.  

There could be several possible explanations for the systematically larger $\esnth$ than $\esnpsh$ and $\esnksh$.
First, it could be due to a systematic error in deriving 
the global parameters from observation;  
the thermal energies could have been systematically overestimated 
or the shell momenta and kinetic energies could have been systematically underestimated.
Thermal energies are derived by assuming that 
hot gas with a pressure determined from X-ray spectral analysis  
is filling the entire volume of an SNR.
Considering that the X-ray parameters are often obtained 
from the analysis of X-ray bright regions in SNRs,
this poses a large uncertainty in the derived thermal energy. 
If we had used the actual volume bright in X-rays, 
$\esnth$ could have been considerably smaller.
For G54.4$-$0.3 that we analyzed 
in Appendix \ref{sec:appendixB1}, 
for example, the X-ray emission is filling an elliptical area of 
$38'\times 22'$, so that the 
thermal energy would be about 40\% 
of $E_{\rm th}$ in Table \ref{tbl-5} 
derived by assuming that the hot gas is filling the entire SNR.
For IC 443, it has been proposed that the X-ray emitting plasma is confined to 
a thin shell \citep{troja06}, the volume of which is only a small 
fraction of the total volume of the SNR as should be the thermal energy. 
Small thermal energy would be more consistent with the observed 
momentum and kinetic energy of an SNR, 
but what fills the rest of the SNR volume needs to be explained. 
One possibility (M. Li et al, in preparation) is that the pressure of cosmic rays becomes
important just inside the shell,  helping to displace the hot gas inwards.  The cosmic rays can be produced either in situ by the diffusive shock acceleration of thermal particles or by the compression/reacceleration of pre-existing cosmic rays. For the middle-aged SNRs interacting with dense MCs such as W44 and IC 443, it has been shown that the latter is sufficient to explain the radio and $\gamma$-ray emission of these SNRs \citep{uchiyama10, lee15}. In either case,  radial expansion of gas in the SNR interior initially compresses and confines cosmic rays within the shell, they can subsequently move downstream into the SNR interior if the thermal pressure there drops below that of the shell. 

Alternatively, the momentum and kinetic energy of the \schi\ shells could have been   
systematically underestimated.  
As was explained in \S~\ref{sec:HI}, 
the global parameters of the \schi\ shells are obtained 
from the thin-shell analysis of small, uncontaminated portion(s) of \schi\ 21 cm emission spectra.
For some SNRs (W44, G54.4$-$0.3, HB21), we assumed that the shell is thin 
and expanding uniformly, 
so that the shell mass per unit LOS velocity is constant and the 
{\em intrinsic} \schi\ 21 cm profile is a rectangle.
(The observed profile appears a flat-topped Gaussian because 
of the large turbulence velocity assumed, e.g., see Figure~\ref{fig:HB21}.)
If there were more mass at low expansion velocities,  
the total mass and therefore the momentum and kinetic energy 
could have been underestimated. 
Since the expansion velocity of this low-velocity gas is small, however, 
its contribution to the derived kinetic energy (and momentum) might not be large 
(see \S~\ref{sec:diss_error}, however). 
The $\esnth$ substantially larger than the canonical SN explosion energy ($1\times 10^{51}$~erg)
in these SNRs also suggests that  the main reason for the inconsistency is probably not  
the underestimation of the shell kinetic energy, although  
careful analysis of SNR simulations in a realistic environment with mock observations 
is necessary to quantify the potential bias in the derivation of shell kinetic energy. 

Another, probably more plausible, explanation for the systematic trend in Figure~\ref{fig:snenergy}
might be that, in deriving SN explosion energy from the global parameters, 
it is difficult to take into  account the complex environments of SNRs, and this is 
compounded by the parameter  sensitivity of Equation \ref{eq:eth_asym} 
(and to a lesser extent Equation \ref{eq:eksh_asym}).
As we already seen in many observed SNRs (e.g., Figure~\ref{fig:w44-ic443}) 
and simulated SNRs (e.g., Figure~\ref{fig:snapshot}), 
the real ISM provides a very complex environment to SNRs. 
In this case, 
the background medium density parameter $\nbg$ in describing SNR evolution is not well defined. 
As demonstrated in 3D simulations considering inhomogeneous background medium 
(e.g., Figure~\ref{fig:sims}; see also \citealt{kim15,cho15,li15,martizzi15,walch15,iffrig15,zhang19}), 
fortunately, the evolutionary tracks of integrated quantities such as total momentum 
and kinetic energy in the radiative SNRs (i.e., after the shell formation) 
are not sensitive to the complexity of the background medium. 
In particular, the total momentum does not evolve in time after reaching 
a terminal value. 
The model uncertainty due to the ISM inhomogeneity is less than a factor of two. 
However, one single background medium density $\nbg$ does not seem be  
applicable to both $\esnth$ and $\esnksh$.
In our TS models, where the ambient medium has a large-scale non-uniformity,  
thermal energy is considerably smaller than that 
of the uniform medium case with the same mean density (Figure~\ref{fig:sims}(e)).   
This also happens for some SNRs in TI models (see \S~\ref{sec:diss_error}). 
On the other hand, the observed thermal energies of the SNRs appear 
to be generally larger than those of the uniform medium cases 
with the same mean densities (Figure~\ref{fig:sims}(f)).
For the two prototypical SNRs interacting with MCs, W44 and IC 443, 
for example, Table \ref{tbl-5} shows that $\esnth/\esnksh$=6--8 
when we adopt the mean ambient density 
$\nbg$(\schi+H$_2$)(=50--70 cm$^{-3}$).
It is, however, clear that this density is much higher than 
the density of the material filling most of the 
volume of the ISM where the SN blast wave propagates.  
For example, the ambient density 
derived from an analysis of the X-ray surface brightness profile of W44 
is $\sim 3$~cm$^{-3}$ \citep{harrus97}.
An overestimated ambient density would predict 
a smaller $\eth$ or a larger $\esnth$  
to match the observed $\eth$ (see Eq. \ref{eq:eth_asym}).
On the other hand, Eq. \ref{eq:eth_asym} is based on 1D simulations. 
In real SNRs in inhomogeneous/non-uniform medium,
the cooling could be significantly enhanced due to the mixing and diffusion
between the engulfed dense material and the hot plasma,
in which case the thermal energy of real SNRs  
might be smaller than $\eth$ predicted from 1D simulations.   
Therefore, it is not obvious if Equation \ref{eq:eth_asym} 
with $\nbg$(\schi+H$_2$) would overpredict or underpredict  
$\esnth$ for SNRs in complex environments (see also \S~\ref{sec:diss_error}). 
High-resolution simulations of SNRs in realistic environments including 
the complex physics at the hot/cool interface are needed in oder to understand the 
environment dependency of the evolution of thermal energy. 

It is worthwhile to point out that $\esnpsh$ and $\esnksh$ in Table~\ref{tbl-5} were derived by assuming that 
the observed momentum and kinetic energy in \schi{}+H$_2$ 
dominate the total budget of momentum and kinetic energy of radiative SNRs, respectively.
This might be a valid assumption for the majority of the SNRs studied in this paper.
But it is still possible that some components of the momentum and kinetic energy are 
undetected (see also \S~\ref{sec:diss_error}). 
For example, the morphology of the SNR G54.4$-$0.3 suggests that the remnant is 
likely to be interacting with large MCs \citep{junkes92a,junkes92b,ranasinghe17}, but 
only a small amount of shocked molecular gas has been detected.  
On the other hand, if an SN explodes in an environment with hot gas filling most of the volume \citep{li15}, 
the observed momentum of the atomic shell can significantly underestimate the 
total injected momentum because no radiative shell forms in the hot gas. 
For a more rigorous comparison of theory and observation, 
it needs to be investigated from numerical simulations 
how the momentum and kinetic and thermal energies are distributed in different phases of the ISM 
in different environments.  

\subsection{Uncertainties in the Thin Shell Analysis}\label{sec:diss_error}

The uncertainties in the derived global parameters of the SNRs are dominated by 
systematic uncertainties in analyzing the observed data. The measurement errors are small.
Our basic assumption in the analysis was that the SNR is composed of a  
fast-expanding, spherical \schi\ shell and hot gas filling the interior.
The shell is assumed to be geometrically `thin' and the 
expansion velocity of the shell material is assumed to be 
either constant or varying radially depending  
on the observed properties of the \schi\ 21 cm emission (see \S~\ref{sec:HI}).
The hot gas is assumed to have a constant pressure, 
although its density and temperature may not be uniform.
We further assumed that all the 
SNR momentum resides in the shell 
(and possibly in shocked molecular gas in addition).  
The errors in the global parameters obtained from 
this thin-shell analysis might depend on the complexity of the environment,
and it is difficult, if not impossible, to quantify the errors for individual SNRs.
By comparing these observation-based global parameters 
with those obtained from the same analysis of mock observations from simulations of SNRs 
in realistic environments, we can in principle 
infer the environments of individual SNRs and/or we can validate theoretical SNR models.
In this work, however, we compared  the observation-based global parameters 
with the {\em total} momentum and {\em total} kinetic/thermal energy 
obtained from numerical simulations. 
We took this approach as the most practical for an initial study 
comparing theory and  observations.
Given the limitations of our approach, however, we consider it worthwhile in this section  to
estimate the uncertainties/errors in the comparison 
arising from applying the thin-shell analysis 
by using one of the TI model SNRs (KO15, see also \S~\ref{sec:comp_sims}). 
The simulation was not intended to be compared in detail to 
any real SNR, but the result will be useful in understanding  
the systematic uncertainties in our analysis.

The simulated SNR is one 
out of the ten realizations
with a mean density of the background medium 
$\nbarc=10$~cm$^{-3}$ (purple lines in Figure~\ref{fig:sims}; S2P-n10 model in Table 2 of KO15).   
The explosion energy was $E_{\rm SN} = 10^{51}$~erg.
We consider the background medium with high ambient density  
that is close to the environment of the middle aged SNRs 
such as W44 where we see a fast expanding \schi\ shell and also 
the shocked molecular gas (see below; see also \citealt{zhang19}).
The background medium consists of two distinct components;     
WNM and CNM with a mean density of 1.5 cm$^{-3}$ and 110 cm$^{-3}$, 
respectively (Figure~\ref{fig:ssnr_structure}). 
The WNM is filling most of the volume with the volume 
filling factor $f_{V, {\rm WNM}}=0.88$,
while the CNM contributes most (83\%) of the mass.  
For convenience, the velocity of the background medium was set to 
zero, so that the SNR material can be easily distinguishable 
by selecting either hot ($T\ge 2\times 10^4$~K) or 
dynamically-perturbed ($|v| \ge 1$~\kms) gas.
We separate the SNR material into different components 
using temperature cuts; neutral ($T < 2\times 10^4$ K), 
ionized ($2\times 10^4$~ K $\le T \le$ $5\times 10^5$~K), 
and X-ray-emitting hot gas ($T > 5 \times 10^5$~K). 
The neutral includes potential molecular gas.
The shell formation time of the WNM is $3.5\times 10^4$~yr (see Eq. \ref{eq:tsf}), 
so we have chosen 
a snapshot at $5\times 10^4$~yr
for our analysis.
Figure~\ref{fig:ssnr_structure} shows the spatial distribution of the 
neutral (in panels (b) and (c)) and hot (in panel (d))
components as well as that of the initial ambient medium (in panel (a)). 
The neutral is further 
divided into two components, (b) slow ($v_r<50$~\kms) and (c) fast ($v_r>50$~\kms),
based on their expansion velocity (see below).
The SNR has an ellipsoidal shape with radial distance 
from the explosion center to the boundary ranging from 15 pc to 21 pc. 
The volume-averaged mean radius is $\rsvol=16.5$ pc while the geometrical mean  
radius of the SNR projected on the sky ($x$-$y$ plane) is 17.8 pc.

The top frame in Figure \ref{fig:ssnr_mass} 
shows the mass distribution of the neutral component of the simulated 
SNR in radial velocity, i.e., $dM/dv_r$ where $v_r$ is radial velocity. 
It shows that $dM/dv_r$ decreases continuously with $v_r$, i.e.,   
there is more mass at lower velocities, which is very different from 
what we expect for a thin shell expanding at a constant speed.
This is because the background medium is 
composed of two distinct components, each of which has a range of densities. 
Most of the mass at high and low velocities are from    
the shocked WNM and the shocked CNM, respectively.
The break in the slope of $dM/dv_r$ at $v_r\sim 50$~\kms\ suggests
that the predominance of the two components switches across this velocity. 
This is more clearly seen in
the momentum distribution $dp/dv_r$ (Figure \ref{fig:ssnr_mass} middle frame), where 
we see that $dp/dv_r$ has a Gaussian-like distribution 
at $v_r \simlt 50$~\kms, while it decreases linearly with $v_r$ at higher velocities.
For convenience, we divide the neutral SNR gas into 
two components: the `slow' component expanding at $v_r < 50$~\kms\ 
and the `fast' component expanding at $v_r \ge 50$~\kms.   
We note that the hydrogen mass of the fast component is 500~\Msun, 
which is comparable to the swept up mass of the WNM 
($\approx (4\pi/3)f_{V,{\rm WNM}} \rsvol^3 \approx 590$~\Msun). 
Figure \ref{fig:ssnr_structure} 
shows the spatial distribution of the slow and fast components.
As expected, the fast component is dominated by the 
expanding shell in the WNM with some additional mass in the interior,
while the slow component is mostly the shocked dense CNM in the interior.
One thing to notice is that the slow component is surrounding the explosion center.
This is because the intial gas distribution is somewhat artificial 
in a sense that the CNM and WNM are isotropically distributed with respect to the SN,
and their mass and volume fractions are solely set by thermal instability without realistic
considerations of turbulence and magnetic fields in the ISM.
The momenta of the two components are comparable while 
the kinetic energy of the fast component is three times greater than 
that of the slow component.  
Table \ref{tbl-6} summarizes the physical parameters of the 
individual components of the SNR as well as those of the entire SNR. 

We adopted $z$ axis as the LOS, and Figure~\ref{fig:ssnr_mass} shows 
 the mass distribution in the LOS velocity $dM/d\vlos$.  
The $dM/d\vlos$ is also composed of two distinct distributions corresponding to 
the slow (solid line) and fast (grey filled area) components. 
The fast component appears as a broad wing 
that extends to high LOS velocities, and 
the mass at $|\vlos|>50$~\kms\ is entirely due to the fast component.
Before we perform an \schi\ 21 cm line analysis of the simulated SNR,
it is worthwhile to consider what we would or would not see in \schi\ 21 cm emission. 
The \schi\ 21 cm emission that has been detected in the SNRs in Table~\ref{tbl-1} 
are all at the highest LOS velocities (e.g., see Figure \ref{fig:HB21}), 
which corresponds to the broad wing due to the fast neutral component in Figure~\ref{fig:ssnr_mass}.  
With turbulence, the slow neutral component with large expansion velocities can 
also have LOS velocities higher than 50~\kms, but its contribution to the detected 
\schi\ emission at the highest LOS velocities will be negligible. 
{\em The \schi\ 21 cm line analysis, therefore, provides the parameters  
of the fast component or the expanding shell in the WNM, 
but it does not provide any information about the slow neutral component or the 
slowly expanding dense material in the interior.}

We have produced a synthetic \schi\ 21 cm line profile of the simulated SNR as in \citet{kim14}. 
We set the spin temperature of the neutral hydrogen ($T<2\times10^4$ K)
equal to the gas kinetic temperature, assuming efficient excitations by
collisions in the CNM and Ly$\alpha$ resonant scattering in the WNM \citep{seon20}.
Note that the \schi\ 21 cm line emission from the shocked SNR \schi\ gas 
is usually optically thin, so that the line intensity is just proportional to the column density 
at high velocities and the details of the excitation and/or radiative transfer 
are not an issue in deriving the \schi\ mass. 
Since we make the ambient medium static in the simulation, we add
a turbulent velocity field to $\vlos$ to mimic random motions in the diffuse \schi.
We generate a Gaussian random velocity field with a power-law slope of 
$-2$ in the wavenumber space and the rms amplitude is set to 10 \kms.
We also add a white noise with an rms amplitude of 0.03 K to the brightness temperature
for an instrumental noise. 

The synthetic \schi\ 21 cm line spectrum of the SNR 
is shown in the top frame of Figure \ref{fig:ssnr_hi} where the dotted line is a  
background spectrum obtained from an annular ring surrounding the SNR.
The middle frame shows the background-subtracted \schi\ spectrum 
where we see the excess emission associated with the SNR 
at velocities higher than $\pm30$~\kms. 
In real SNRs, only the emission at the highest velocities are seen 
because the emission from the foreground/background \schi\ gas is very broad due 
to the Galactic rotation.
We assume that either only the spectrum at $|\vlos| \ge 70$~\kms\ 
or at $|\vlos|\ge 90$~\kms\ is visible, and perform a least squares fitting 
to obtain the expansion velocity and the hydrogen column 
density, which are used to derive the age, mass, momentum, and kinetic energy of 
the shell (see \S~2.1). For the radius of the SNR, we use the  
geometrical mean radius (17.8 pc) as it appears on the sky.
For the expanding shell, we adopt the simplest thin shell model
where the shell material is expanding at a constant speed. 
But the shell is assumed to be turbulent with a large 
($\Delta v_{\rm FWHM}=50$~\kms) dispersion in the LOS velocity, so that the model   
\schi\ 21 cm spectrum appears as a flat-topped Gaussian.
This velocity width is from the observation 
of \schi\ clumps in the expanding SNR shells 
and it is what had been used for the fit in previous studies \citep[e.g.,][]{park13}.
The details of the fitting procedure may be found 
in \cite{park13} (see also Appendix A).
The bottom frame of Figure~\ref{fig:ssnr_hi} 
is zoom-in of the middle frame and
shows the best fit profiles, and 
Table \ref{tbl-7} summarizes the derived shell parameters. 

Table \ref{tbl-7} shows that the global parameters 
derived from the \schi\ 21 cm line analysis agree well with those of the 
fast component of the simulated SNR.
When using only the \schi\ profile at the highest LOS velocities ($|\vlos|\ge 90$~\kms), 
the mass is substantially ($\simgt 30$\%) underestimated. 
On the other hand, the derived expansion velocity is larger than 
the mass-weighted radial velocity of the SNR shell, so that 
the momentum and kinetic energy agree with those of the simulated SNR 
within $\sim 10$\%. The age also agrees with the age of the simulated SNR.
When the \schi\ profile at $|\vlos|\ge 70$~\kms\ is used, the mass becomes larger, 
and the momentum and kinetic energy are overestimated by $\sim 20$\%.
{\em So, for the simulated SNR analyzed in this section, 
the errors in the momentum and kinetic energy of the fast neutral component of the SNR  
arising from the thin-shell analysis are small ($\simlt 20$\%).}
There is slowly-expanding neutral material inside SNR, which is not included in the analysis.  
Its momentum is slightly larger than that of the expanding shell while its 
kinetic energy is much less (Table~\ref{tbl-6}).
This slow neutral component is not likely to be 
traced by the \schi\ 21 cm line emission in the observations of real SNRs 
because of the background/foreground contamination.  
Hence, the comparison cannot be made. 
If the slow neutral component is molecular,  however,  
it can be detected in molecular emission lines and can be 
included in the comparison as in the SNRs W44 and IC 443.

We can also estimate the error in 
the thermal energy arising from the thin-shell analysis.
In the simulated SNR, the volume filling factor of the hot gas  
is 53\%. The rest is filled with the neutral (28\%) and the ionzied (19\%).
Hence, if we derive thermal energy assuming that 
the hot gas is filling the entire SNR having a spherical volume of 
radius $\rshell$ (=17.8 pc), which is larger than the volume-averaged radius (16.5 pc), 
the thermal energy would be overestimated by a factor of 2.4. 

If we naively accept the result of the above analysis, the momentum of the
radiative `shell' in Table~\ref{tbl-4} could be underestimated by as much as
a factor of ($\pshell$(\schi)+$\pshell$(H$_2$))/(2$\pshell$(\schi)+$\pshell($H$_2$)) 
because the slow component is not included. 
If the shocked molecular gas 
fully accounts for
the slow component, 
however, the error is  small ($\simlt 20$\%).
The error in the kinetic energy due to the missing slow component 
is small ($\simlt 30$\%). 
Thermal energy could be overestimated by a factor of 2--3
due to the geometrical uncertainty.
The error in the age of the shell appears to be small.

Finally, we estimate the error in the derived explosion energy from the
``observed'' global parameters of the simulated SNR. 
Table~\ref{tbl-8} shows $\esn$ obtained from the parameters of 
the $|\vlos|\ge 70$~\kms\ case in Table~\ref{tbl-7}.
(The results are essentially the same for $|\vlos|\ge 90$~\kms.)
With the background density $\nbarc=0.90\pcc$,   
we obtain $\esnpsh \approx \esnksh \approx 0.23\times10^{51}$ erg, i.e., 
the derived explosion energy from the momentum 
and kinetic energy of the shell is consistently lower by about a factor of 4.
We also show the results in a hypothetical case 
when the slow component ($|v_r|<50$~\kms; Table~\ref{tbl-6}) is 
observed, e.g., in molecular lines, 
and its mass, momentum, and kinetic energy are included in the analysis. 
(Note that this corresponds to the \schi+H$_2$ case in Table~\ref{tbl-5}.)
In principle, $\esn$ should be close to 1 in the latter case, but because 
the analytic evolution tracks are generally above the evolution tracks of the TI models
(see Figure~\ref{fig:sims}), $\esnpsh$ and $\esnksh$ are underestimated by a factor of 2.  
Therefore, actual bias introduced by the missing slow component is a factor of 2
underestimation in the derived explosion energy.
For $\esnth$, we use $\eth=4.2\times 10^{49}$~erg, which is 2.4 
times the thermal energy of the hot gas (see the above paragraph). 
We obtain $\esnth=0.17\times 10^{51}$~erg for $\nbg=0.90\pcc$ and $\esnth=0.63\times 10^{51}$~erg for $\nbg=6.7\pcc$. 
Note that $\esnth$ is underestimated, although $\eth$ has been overestimated. 
This is again because the analytic evolution track is 
above the evolution tracks of the TI models (Figure~\ref{fig:sims});  thermal energy of 
the simulated SNR drops rapidly after the shell formation, so that  
$\esnth$ derived from the analytic evolution track is underestimated by a factor of 3.  
Therefore, actual bias introduced by the geometrical uncertainty 
is a factor of 2 ($=3\times 0.63$) {\em overestimation} in the derived explosion energy.
If the slow component is not included ($\nbg=0.90\pcc$),
$\esnth$ is underestimated by a factor of 2 ($=1/(3\times0.17)$) because of 
the low background density (see Eq. \ref{eq:eth_asym}).
Hence, aside from the uncertainties associated with the analytic evolution tracks, 
$\esnth$, as well as $\esnpsh$ and $\esnksh$, in Table~\ref{tbl-5} 
would be underestimated by a factor of 2 due to the missing slow component.
For $\esn$ obtained with the background density of \schi+H$_2$, 
if the shocked molecular gas corresponds to the slow component, 
$\esnth$ would be overestimated by a factor of 2 due to geometrical 
uncertainty, while the errors in $\esnpsh$ and $\esnksh$ might be small.

How general is the above result? 
It has been shown that {\em total} momentum/kinetic energy of the SNRs 
in simulations after shell formation differs by less than a factor of 2 
in a wide variety of background conditions (see \S~3.2). 
The relative distribution in different phases of the ISM, however, 
could be very different depending on the environment as well as the age of the SNR. 
On the other hand, it does not seem unreasonable to expect that the 
SNRs of similar ages with fast expanding \schi\ shells detected have  
similar environments. For example, if the ambient medium is rarefied 
the \schi\ shell will not be formed until the SNR becomes very old 
(e.g., $\simgt 10^5$~yr when $\simlt 0.1$~cm$^{-3}$; Eq. \ref{eq:tsf}).
Or, if the ambient medium is dense, the \schi\ shell will be detected only 
for relatively young SNRs because the expansion velocity of the \schi\ shell 
will drop below the detection limit in the early stages of their evolution. 
Hence, the parameter space to be searched is likely to be restricted. 
We consider that 
the thermal energy content of hot gas might be important in exploring the parameter space.
As we mentioned in \S~3.2, the thermal energy evolution of radiative SNRs
is sensitive to the structure of the background medium. 
Indeed the thermal energy of the hot gas in the simulated SNR 
($1.7 \times 10^{49}$ erg) is more than a factor of 4 smaller than 
the thermal energy predicted from 
Equation \ref{eq:eth_asym} which is $7.4\times 10^{49}$~erg.
This suggests that the discrepancy in thermal energies found in \S~3.2 
could be environmental, although, as we have seen in the simulated SNR, 
the thermal energies obtained by assuming a spherical 
volume filled with hot gas 
could have been overestimated by a factor of 2--3. 
Owing to many simplifications made to obtain the initial ambient medium
and some missing physics in the simulated SNR as well as the simplifications 
in creating the mock observation (see \S~\ref{sec:diss_physics}), 
the current analysis should be considered 
as one case study to understand the potential errors 
in comparing the observation-based global parameters with simulations. 
It will be interesting to look into the momentum contribution from 
the slow neutral component and the thermal energy issue 
with realistic simulations and in a wider parameter space in future works.

\subsection{Missing Physics in Numerical Models}\label{sec:diss_physics}

There are a few physics elements missing in the three-dimensional simulations presented here, 
including thermal conduction and cosmic rays as well as
magnetic fields. The role of the interstellar magnetic fields has been briefly
investigated in KO15, demonstrating that magnetic fields that are stronger than
$B\sim 7\ \mu{\rm G}$ can alter the shape of an SNR and reduce the late time
momentum slightly as the shell expands back to the interior due to strong
magnetic pressure built in the shell. Yet, the reduction is slight, well within
the uncertainty we consider here.

Thermal conduction transfers energy from the interior to the shell, which is
compensated by energy delivered by flow from the shell due to evaporation. This
effect is actually included in the one-dimensional simulations presented in
this paper, using the same framework developed for superbubble evolution driven
by multiple SNe in \citet{el-badry19}. For a single SN, the evolution is almost
identical without thermal conduction. 

In the clumpy ISM, the role of thermal conduction can be enhanced and more generally termed
as the effect of ``mass loading''. As a result of interaction between
blastwaves and embedded cold clouds, cold clouds are evaporated by thermal
conduction \citep[e.g.,][]{cowie77} and shredded by hydrodynamic instabilities
\citep[e.g.,][]{klein94}. Then, the interior hot remnant loads more mass and
cools more rapidly, altering the overall evolution of the integrated properties of
SNRs. The ``mass loading'' effect is in part modeled in KO15 (and the TI models
presented in Figure~\ref{fig:sims}) as it explicitly simulates blast wave
expansion within a clumpy medium, while the resolution requirement to resolve
cloud crushing is more stringent \citep[e.g.,][]{schneider17}. 
\citet{slavin17} and \citet{zhanggao19} conducted a set of 
2D and 3D simulations for SNR expansion in cloudy medium 
with thermal conduction and modeled X-ray emission to explain the 
centrally-peaked X-ray in MM SNRs. Since radiative cooling is ignored 
in their simulations (adiabatic simulations), unfortunately, a direct link between 
X-ray emission and remaining thermal energy is not possible. 
Although the evolution is limited to a time shorter than the shell formation 
time of the background medium, blastwaves propagate into clouds as well as 
gas in the interface and wakes would have cooled. 
\citet{zhang19} modeled a radiative SNR in a turbulent medium (without conduction) 
and showed that total energy and momentum evolution is not very sensitive 
to the Mach number of the medium. 
The interior X-ray emission has brightened in the turbulent medium, 
but not as bright as its adiabatic counterpart. 

Besides full 3D
simulations with explicit treatment of the inhomogeneous ISM, the effect of
``mass loading'' has been investigated using similarity solutions
\citep[e.g.,][]{mckee77,white91} and one-dimensional simulations
\citep[e.g.,][]{cowie81,pittard19}. Very recently, \citet{pittard19} suggests
that the final momentum can be reduced significantly by the mass loading from
cold gas. However, even for the case with $f_{\rm
ML}=100$, in which mass loaded by clouds is 100 times larger than the swept-up mass at
the shell formation, the reduction of the final momentum is only 56\%, which
again well within the range of model uncertainty considered here. 
Furthermore, the mass loading generally enhances cooling 
so that the thermal energy of SNRs might be reduced compared to that without the mass loading,  
which would increase the discrepancy between 
the explosion energies inferred from thermal and momentum/kinetic energy. 
3D simulations of a radiative SNR in a inhomogeneous medium 
with thermal conduction at high resolution 
will illuminate the role of mass loading in dynamical evolution 
and observational signatures of MM SNRs.

The last piece of the puzzle in SNR evolution study is the role of cosmic rays. It
is well accepted that blastwave shocks are the main source of energetic
particles \citep[e.g.,][]{bell78,blandford78}. As cosmic ray energy is
$\sim10-20\%$ of kinetic energy of SN ejecta \citep[e.g.,][]{caprioli14,park15}
and hardly radiated away, cosmic ray pressure can be a source of
further acceleration in the pressure-driven snowplow phase \emph{if cosmic rays
are confined preferentially behind the shell}. Recent semi-analytic analysis
suggests that the final momentum deposition can be enhanced by more than a factor of 5
with $E_{\rm CR}\sim 0.1\esn$ \citep{diesing18}. However, the efficiency of cosmic ray momentum
boost depends strongly on the evolving spatial distribution of cosmic rays relative to the gas concentrated in  the shell, which depends on cosmic ray  streaming and diffusion (M. Li et al. 2020 in prep.).
Therefore, the impact of cosmic rays on SNR evolution is uncertain yet and an
active area of research. 

Overall, the theoretical uncertainty from numerical studies in describing the
evolution of momentum and kinetic energy of radiative SNRs is less than a
factor of two for commonly explored physics (e.g., magnetic fields, background
medium inhomogeneity and turbulence, thermal conduction) unless cosmic rays
significantly alter the momentum deposition in the high density medium. This
allows us to derive the explosion energy relatively reliably from momentum and
kinetic energy of radiative SNRs if the observed momentum and kinetic energy are the majority.
As discussed in Section~\ref{sec:diss_error}, uncertainties at the level 
of a factor  $\sim2$--4 can arise from applying the thin, 
spherical shell model together with analytic evolution tracks
to \schi\ emission lines in order to 
estimate momentum and kinetic energy, or infer the explosion energy.
On the other hand, the thermal energy evolution of the radiative SNR is sensitive to the density 
of the surrounding medium, giving rise to a large model uncertainty in thermal energy 
and the derived explosion energy.

\section{Conclusion}\label{sec:conclusion}

Supernova explosions control the evolution of the ISM and galaxies 
by injecting prodigious energy and momentum. 
The energy produces the hot phase of the ISM,
while the momentum drives turbulence and regulates the star formation.
To understand the ISM structure and evolution of galaxies, therefore, 
it is important to `quantify' the SN feedback, particularly 
the momentum transferred to the ISM
and the amount of hot gas created 
during the lifetime of an SNR. 
There have been several 3D numerical simulations 
for such purpose
\citep[e.g., KO15;][]{li15,martizzi15,walch15,iffrig15,zhang19}.
According to these numerical studies,   
the radial momentum and kinetic energy of SNRs  
are insensitive to the density 
structure of the ISM, so that they are only slightly ($\simlt 50\%$) 
different from those in a uniform medium of the same mean density. 
The final momentum injected to the ISM by a single SN explosion is 
$\pshell=(1-5)\times10^5$~\Msun\kms.
Observationally, however, it is very difficult to detect  
radiative shells associated with SNRs 
because of the contamination due to the background/foreground \schi\ emission.
So there are only a limited number of SNRs with radiative expanding shells detected and, even in those SNRs, we are not seeing all of the shell material but only the fastest expanding portion of the shell. 
In previous studies, the parameters 
of the expanding shells have been derived from the thin-shell analysis of  
\schi\ 21 cm emission line data. 
In this study, for seven radiative SNRs with fast expanding \schi\ shells, 
we have inventoried their shell parameters and also their thermal energies and  
the parameters of the shocked molecular gas associated with the SNRs, 
and compared them 
with the results of 1D and 3D numerical simulations in realistic environments.
We also carried out 
a mock observation of a simulated SNR and discussed   
the uncertainties/difficulties in the comparison.
In the following we summarize the main results of this work:

\begin{enumerate}
\item We provide a table (Table~\ref{tbl-1}) summarizing the 
global parameters 
of radiative \schi\ shells in the seven SNRs 
obtained from the thin-shell analysis of \schi\ 21 cm emission line data. 
All parameters 
are from the literature except those of HB 21 which are derived in this paper. 
The momentum, kinetic energy, and the age of the SNRs are in the range of   
$\pshell=(0.5$--$4.5)\times 10^5$ \Msun~\kms, 
$\ekshell=(0.4$--$3.5)\times 10^{50}$ erg, and 
$\tshell=(0.3$--$3.4)\times 10^5$~yr, respectively. 

\item Among the seven SNRs, four SNRs are known to be interacting with MCs (see Table~\ref{tbl-2}).
In W44 and IC 443, the momentum of shocked molecular gas is comparable to or even larger than  
that of the \schi\ shell (1.8 and $0.7\times 10^5$ \Msun~\kms), while its 
kinetic energy is relatively small ($\simlt 0.2\times 10^{50}$ erg).  
In the other SNRs, the shocked molecular gas is either not prominent or has not been detected, 
so that the momentum and kinetic energy of shocked molecular gas might not be significant.

\item Table~\ref{tbl-4} summarizes the global parameters of the seven SNRs.
The total (atomic+molecular) momentum and kinetic energy of the SNRs are in the range of 
$\pshell=$(1.1--4.5)$\times 10^5$ \Msun~\kms\ and  
$\ekshell=$(0.6--3.5)$\times 10^{50}$ erg, respectively. 
The table also lists the thermal energies of the SNRs 
(see also Table~\ref{tbl-3}). 
They are from the literature except those of G54.4$-$0.3 and CTB 80 which are derived in this work.
The thermal energy, assuming that the interior of the 
SNR is filled with hot gas, is in the range of $\eth=(0.6-8)\times 10^{50}$~erg.   

\item 
The observation-based global parameters of SNRs are generally consistent with 
the 1D hydrodynamic simulations. In particular, the momenta of 
seven SNRs are all very close to the expected terminal values 
based on numerical models for the canonical SN explosion 
energy of $10^{51}~{\rm erg}$. 
By comparing with a variety of 3D hydrodynamic simulations with different background medium states, we show that the systematic uncertainty in determining the explosion energy is less than a factor of 2 for momentum and kinetic energy. 
Modulo systematic uncertainties in the global parameters, 
our inferred explosion energy (Table~\ref{tbl-5}) is consistent with $10^{51}{\rm\,erg}$. 
Thermal energy, however, depends strongly on the density structure of the background medium, 
so that a single mean density of the background medium in general cannot be used 
to characterize both thermal and kinetic energies.   

\item
We explored the uncertainties arising from applying the thin-shell analysis 
by using a mock \schi\ observation of
a simulated SNR in the two phase ISM. 
It shows that there could be dense, slowly-expanding 
neutral material unseen in \schi\ emission and hence
not included in the thin-shell analysis unless it is molecular. 
In the simulated SNR, this slow neutral component 
has a substantial volume filling factor and its momentum is comparable to that of the 
fast expanding \schi\ shell.  
The result might depend on the environment as well as the SNR age, 
and it needs to be investigated from numerical simulations how the momentum and
kinetic/thermal energy are distributed in different phases of the ISM in different environments.

\end{enumerate}

\acknowledgements 

\begin{figure}[h]
\begin{center}
\includegraphics[scale=0.65]{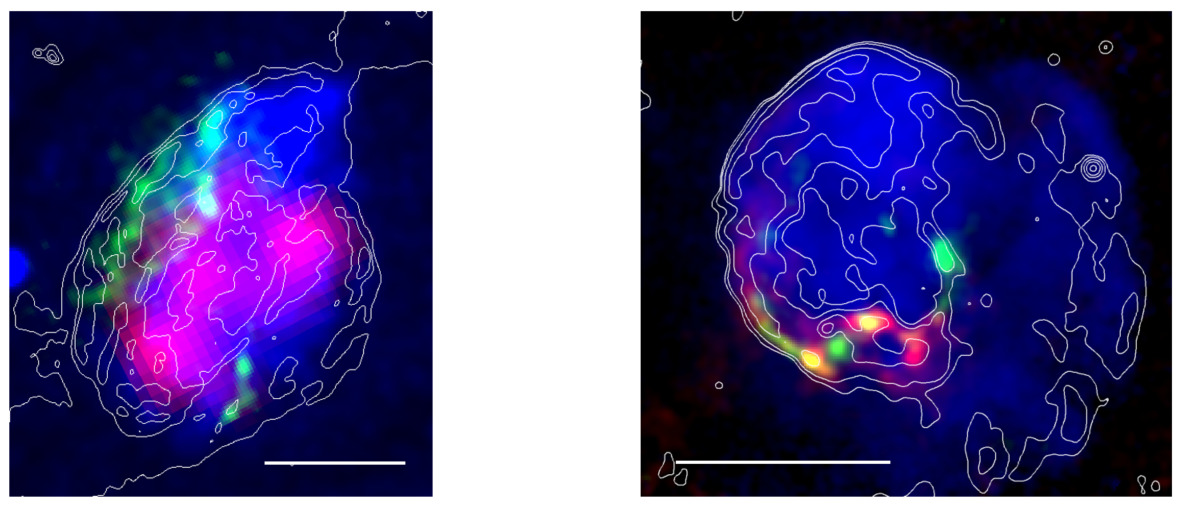}
\caption{Two prototypical SNRs with 
both shocked atomic and shocked molecular gases: W44 (left) and IC 443 (right).
North is up and east is to the left.
Red=shocked atomic gas in \schi\ 21 cm emission 
(W44: \citealt{park13} , IC443: \citealt{lee08}), 
Green=shocked molecular gas in 
HCO$^+$ J=1--0 line (W44:  \citealt{sashida13}, IC443: \citealt{lee12}), 
Blue=shocked hot gas in X-ray from Chandra 
(W44: \citealt{rho94}, IC443: \citealt{asaoka94}), 
Contour=21 cm continuum (W44:  \citealt{giacani97}, IC 443: \citealt{lee08}). 
The scale bar in each frame represents 10 pc at the assumed distance of the SNR 
(see Table~\ref{tbl-1}). 
}\label{fig:w44-ic443}
\end{center}
\end{figure}

\begin{figure}[h]
\begin{center}
\includegraphics[scale=0.8]{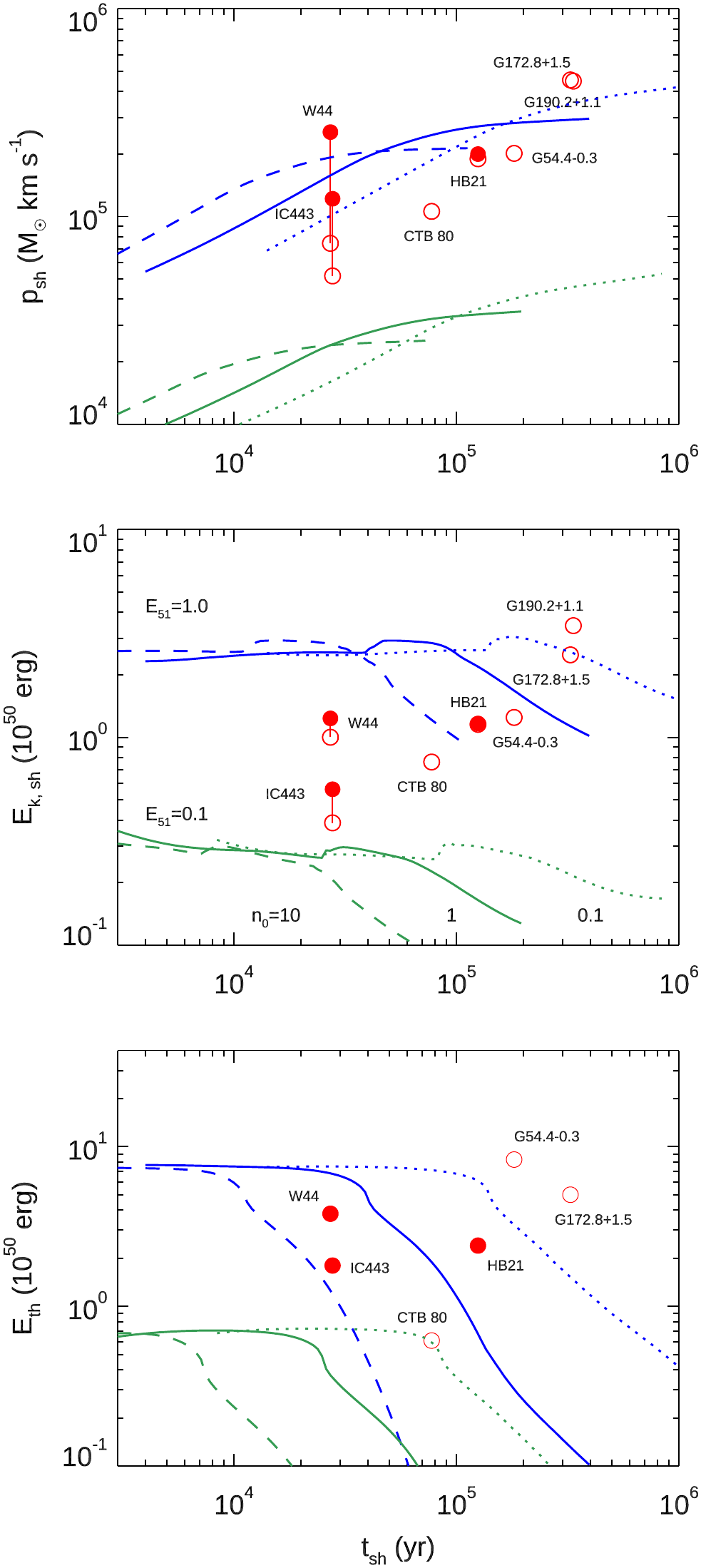}
\caption{Observation-based momentum ($\pshell$), kinetic energy ($\ekshell$), and thermal energy ($\eth$) 
of the SNRs with fast-expanding \schi\ shells. The $x$-axis is the characteristics age of 
the \schi\ shell 
$\tshell = 0.3\rshell/\vshell$. The open circles represent the \schi\ component only, while the filled circles include the contribution from shocked molecular gas. 
For the SNRs marked only with open circles, the contribution from molecular component 
is not significant (see Table~\ref{tbl-2}).  
Note that $\pshell, \ekshell$ and $\eth$ are obtained by assuming that 
the \schi\ shell is thin and spherical and that the hot gas is filling the interior of the SNR 
(see text for more details). 
The formal errors are small and not shown here (see Tables~\ref{tbl-1}--\ref{tbl-3}).
Also shown are theoretical evolutionary tracks obtained from 1-D hydrodynamic simulations of expansion into a uniform medium. 
The blue and green lines are for SN explosion energies $\esn=10^{51}$ and $10^{50}$ erg, respectively,
while the dotted, solid, and dashed lines represent different ambient densities 
($n_0 = (n_{\rm bg}/1~{\rm cm}^{-3}) = 0.1, 1$, and 10).   
}
\label{fig:obs}
\end{center}
\end{figure}

\begin{figure}[h]
\begin{center}
\includegraphics[width=\textwidth]{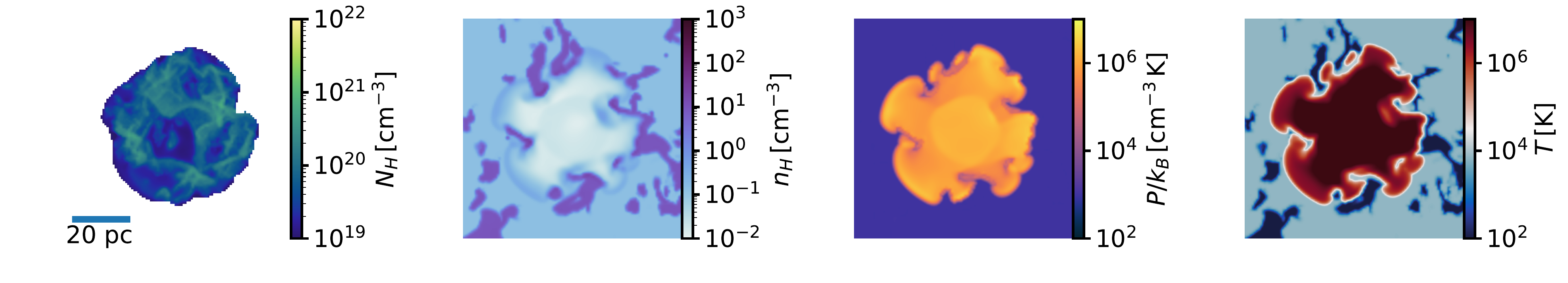}
\includegraphics[width=\textwidth]{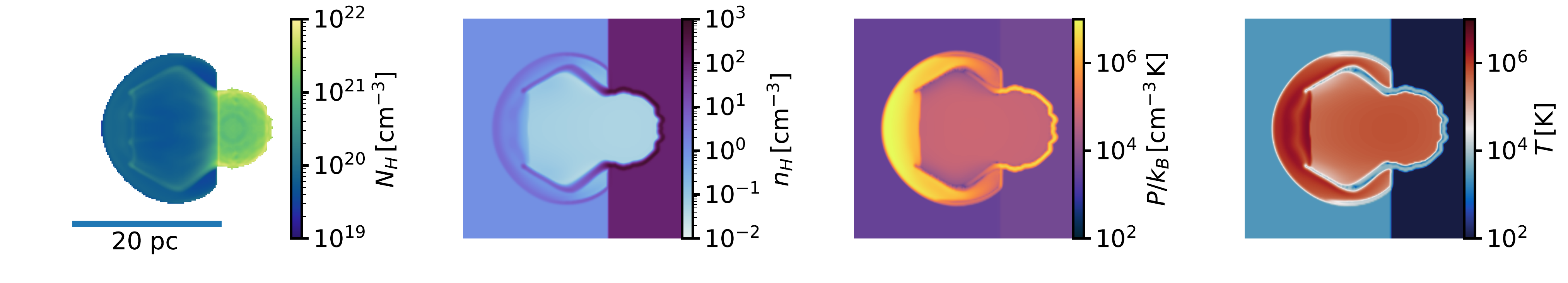}
\caption{From left to right, we show SNR column density and slices of number density, thermal pressure, and temperature. The SNR component is selected by $T>2\times10^4{\rm\,K}$ and $|v|>1{\rm\,km\,s^{-1}}$ (see KO15). 20 pc scale bars are shown on the bottom left corner of the column density panels.
{\bf Top Row:} An example snapshot at $t=0.03{\rm\, Myr}$ of a TI model with {\em mean} hydrogen number density $\bar\nbg=1$~cm$^{-3}$. The two phase medium is produced by running a thermal instability simulation for 200 Myr to reach a nonlinear saturation state. The CNM and WNM have typical hydrogen number density of $\sim9\pcc$ and $\sim0.15\pcc$, respectively.
{\bf Bottom Row:} An example snapshot of a TS model with the explosion depth $h=2{\rm \,pc}$ at $t=0.03{\rm\,Myr}$. The density of two medium is 1 and $100\pcc$.
For references, the shell formation times in uniform media with 
$n_{\rm bg}=0.1$, 1, 10, and 100~cm$^{-3}$ are $\tsf=0.16$, 0.044, 0.012, and 0.0035 Myr. Therefore, blast waves already cool in the CNM of the TI model and the dense medium of the TS model. 
\label{fig:snapshot}}
\end{center}
\end{figure}

\begin{figure}[h]
\begin{center}
\includegraphics[width=0.8\textwidth]{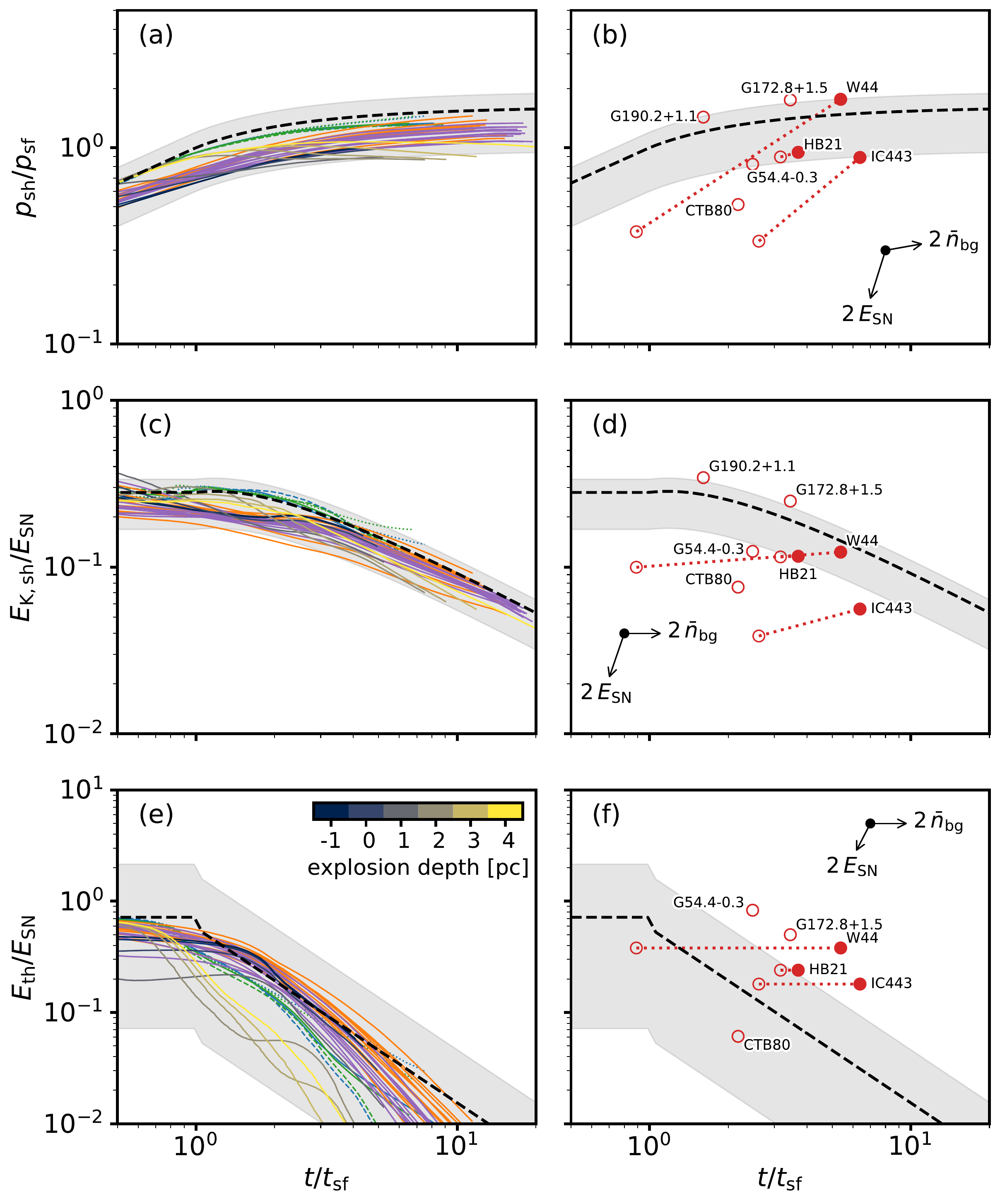}
\caption{
{\bf Left Column:} Evolution of (a) total radial momentum, (c) kinetic energy,
and (e) thermal energy of the simulated SNRs. The properties of SNRs are
normalized by the quantities at the shell formation (i.e., $\tsf$ and
$\psf$; see \S~\ref{sec:snr_prop}) and 
the explosion energy.
The simulated evolution tracks of SNRs in a uniform
medium (blue and green lines same as Figure~\ref{fig:obs}) are almost
congruent for the normalized properties. The analytic model 
(Eqs. \ref{eq:psh_asym} and \ref{eq:eksh_asym}) describing the 1D model is shown as black dashed line. 
The evolution tracks from the TI models are shown as two groups of solid lines
($\nbarc=1$~cm$^{-3}$ for orange and $\nbarc=10$~cm$^{-3}$ for purple),
each of which corresponds 10 realizations of the background medium as a result of thermal instability.
The evolution tracks from the TS models are color coded by the explosion depth from the interface 
toward the denser medium as indicated by the legend in (e). 
Both evolution tracks generally follow the analytic model, especially for momentum and kinetic energy. 
The gray shaded region covers the arbitrarily rescaled analytic models to gauge model uncertainties in describing simulations. The area spans 0.6-1.2$\times$ the model for momentum and kinetic energy and 0.1-3$\times$ the model for thermal energy. 
{\bf Right Column:} Observed SNR properties normalized by the quantities 
at the shell formation using the ambient medium density ($\bar\nbg$ in
Table~\ref{tbl-4}; either \schi\ only or \schi+H$_2$) and the canonical explosion energy
($\esn=10^{51}$ erg). 
The vectors in each panel show the
directions for systematic uncertainties of $\esn$ and $\bar\nbg$ 
adopted for each SNR. The same analytic model curve and area is duplicated for comparison.
\label{fig:sims}}
\end{center}
\end{figure}

\begin{figure}[h]
\begin{center}
\includegraphics[scale=0.8]{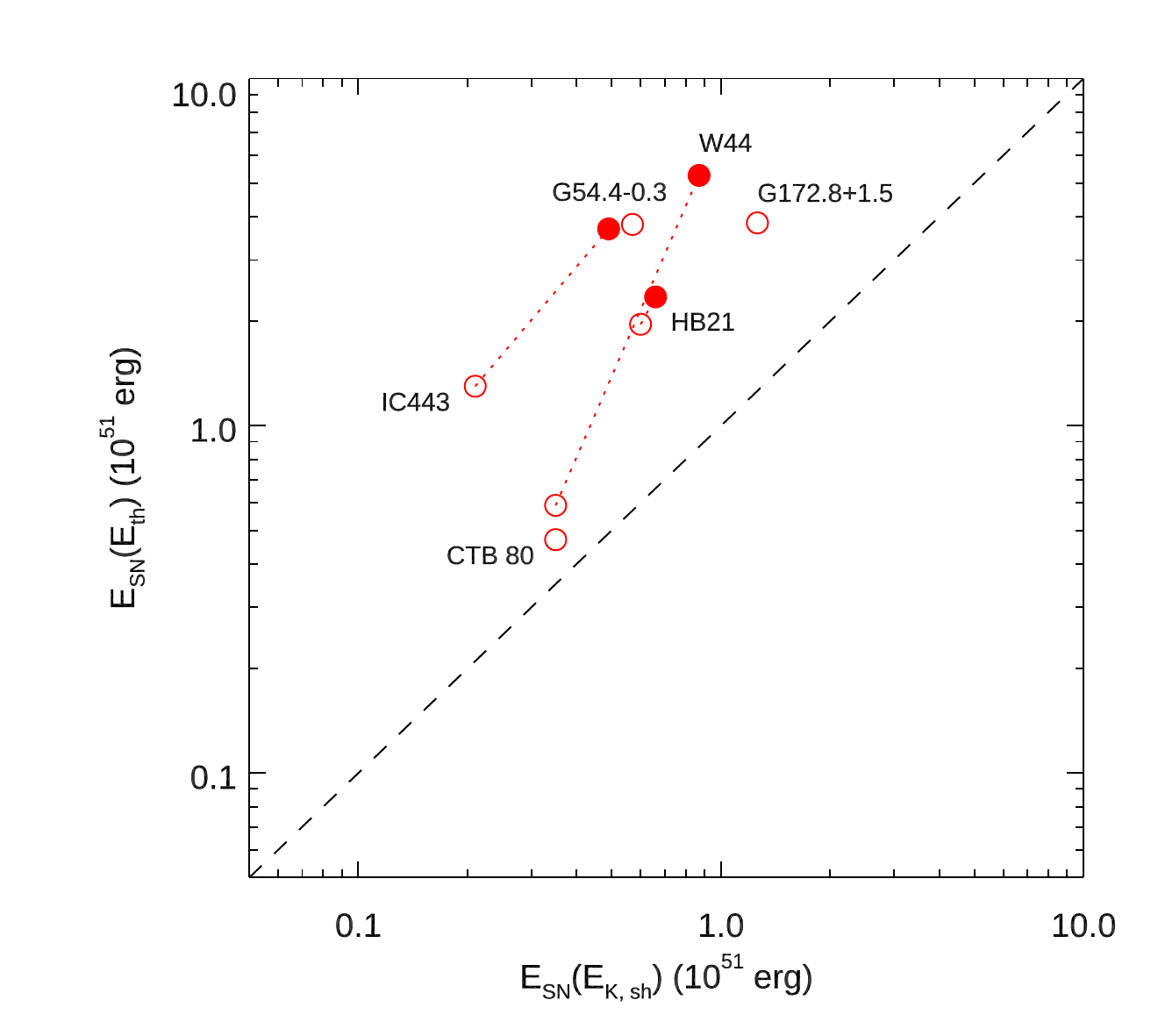}
\caption{Comparison of supernova explosion energies 
calculated from kinetic energy of \schi(+H$_2$) shell ($\esnksh$) and 
thermal energy ($\esnth$). 
The symbols are same as Figure~\ref{fig:sims}, i.e., 
\schi\ only (open circles) and \schi+H$_2$ (filled circles).
}
\label{fig:snenergy}
\end{center}
\end{figure}

\begin{figure}[h]
\begin{center}
\includegraphics[scale=0.4]{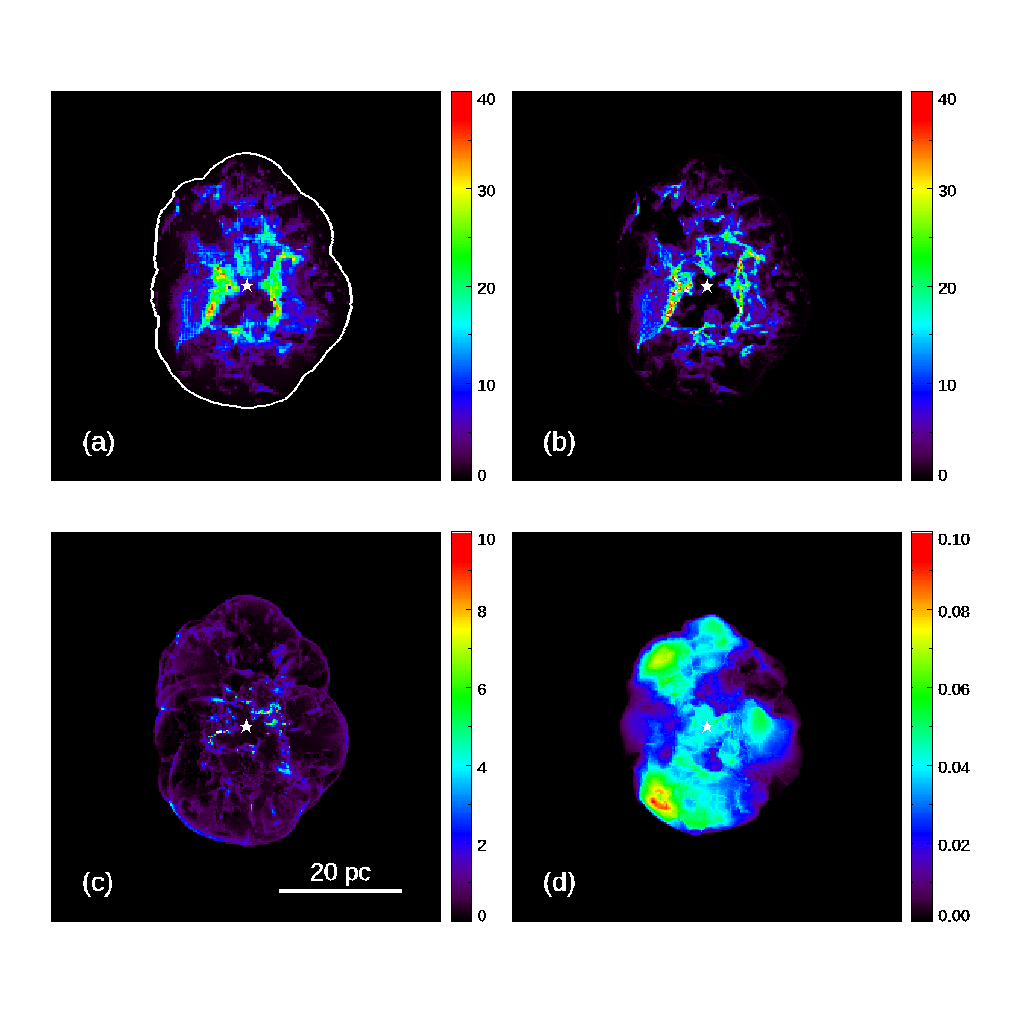}
\caption{Projected hydrogen column densities of initial ambient medium 
and the SNR material: (a) initial ambient medium,  
(b) slow neutral component expanding at $v_r < 50$~\kms, 
(c) fast neutral component expanding at $v_r \ge 50$~\kms, 
and (d) hot component with $T> 5\times 10^5$~K. 
The SNR has been produced in a two phase ISM with $\nbarc=10$~cm$^{-3}$, and 
it is $5\times 10^4$ yrs old (see \S~\ref{sec:diss_error}).
In panel (a), we show the initial column density distribution 
(before the explosion) of material that is later shocked and joins the SNR.  
The white contour represents the projected outer boundary of the SNR.
The scale in each frame is in units of $1\times 10^{20}$~cm$^{-2}$.
}
\label{fig:ssnr_structure}
\end{center}
\end{figure}

\begin{figure}[h]
\begin{center}
\includegraphics[scale=0.8]{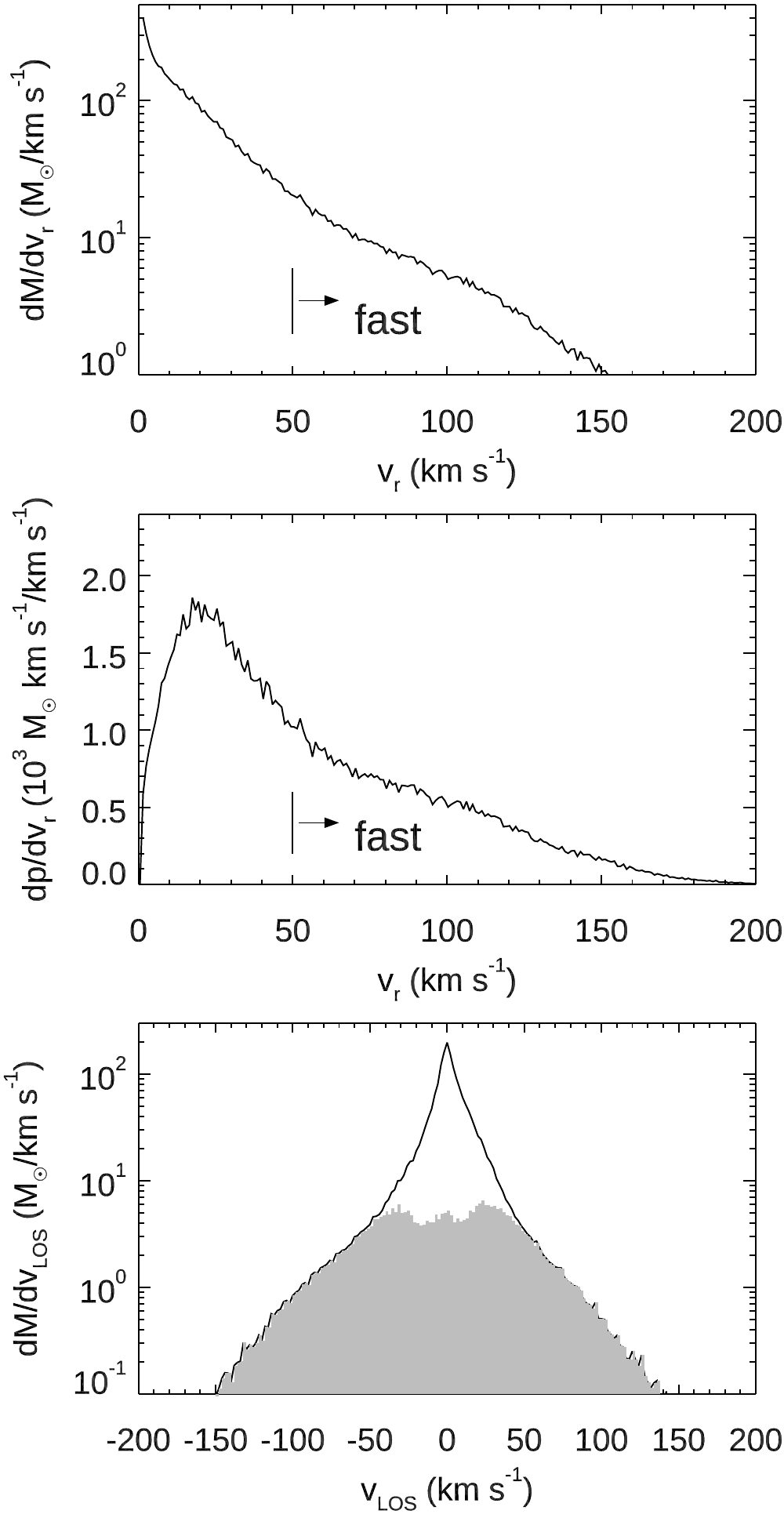}
\caption{ Mass and momentum distribution of neural gas in the simulated SNR in velocity: 
mass distribution in radial velocity (top),  
momentum distribution in radial velocity (middle), and   
mass distribution in LOS velocity (bottom). 
The filled area in the bottom frame shows the distribution of the 
fast ($v_r \ge 50$~\kms) neutral component.
}
\label{fig:ssnr_mass}
\end{center}
\end{figure}

\begin{figure}[h]
\begin{center}
\includegraphics[scale=0.8]{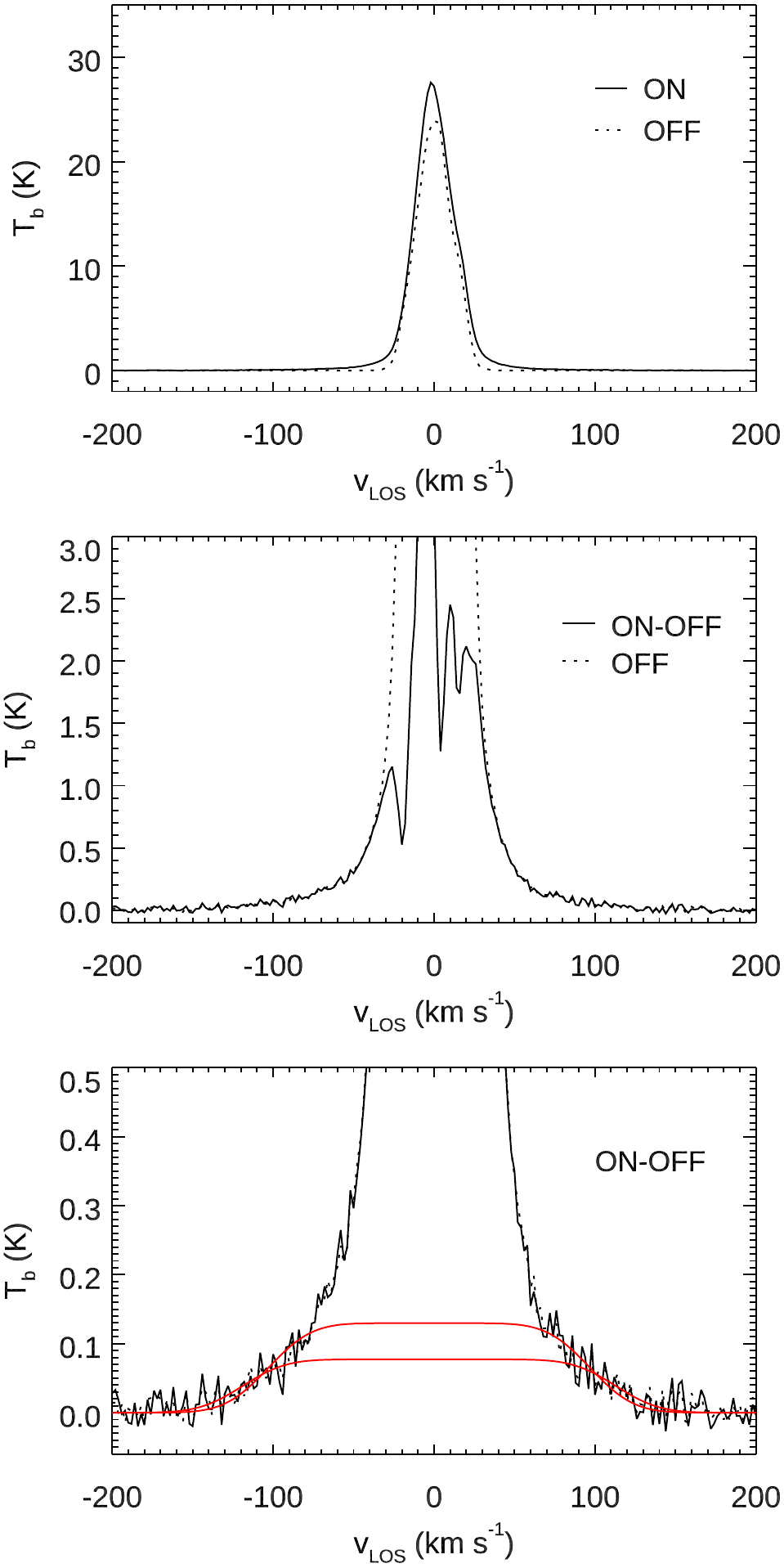}
\caption{{\bf Top:} Source and background \schi\ 21 cm profiles of 
the simulated SNR. The source profile is the average profile inside the 
circular area of radius 20 pc, while   
the background profile is the average profile inside the annulus of 
inner and outer radii of $22$ pc and $24$ pc, respectively.
{\bf Middle:}  Background-subtracted \schi\ spectrum (solid line).  
{\bf Bottom:} A close-up view of the 
background-subtracted \schi\ spectrum and a fit. 
The two red solid lines are the best fits to the profile 
at $|v_{\rm LOS}| > 70$~\kms\ and 90~\kms, respectively (see text).   
}
\label{fig:ssnr_hi}
\end{center}
\end{figure}

We thank the referee, Pat Slane, for his constructive comments 
which helped to improve the paper.
We also wish to thank Tomoharu Oka and Tomoro Sashida for providing their 
HCO$^+$ $J=1$--0 line data used in Figure~\ref{fig:w44-ic443}. 
B.-C.K. gratefully acknowledge the helpful discussions with Chris McKee.
This research was supported by Basic Science 
Research Program through the National Research Foundation of
Korea(NRF) funded by the Ministry of Science, ICT and future Planning (2019R1A2B5B01001994).
The work of C.-G.K. and E.C.O was partly supported by a grant from the Simons Foundation (528307, ECO)
and NASA (ATP NNX17AG26G).
\vspace{5mm}

\software{mpfit \citep{markwardt09}, Athena++ \citep{stone20} }

\bibliography{ref}

\clearpage

\newpage
\clearpage

\newpage
\clearpage

\newpage
\clearpage

\newpage
\clearpage

\newpage
\clearpage

\newpage
\clearpage

\newpage
\clearpage

\newpage
\clearpage

\begin{deluxetable}{l l c c D c D c D D c}
\tabletypesize{\scriptsize}
\tablecaption{Parameters of Radiative HI Shells Associated with SNRs\label{tbl-1}}
\tablewidth{0pt}
\tablehead{
\colhead{G-name} &  \colhead{Name} &   \colhead{d} &  \colhead{$v_0$} &   \multicolumn2c{$\rshell$} &      
\colhead{$\vshell$} & \multicolumn2c{$\tshell$} & \colhead{$\mshell$(\schi)} & \multicolumn2c{$\pshell$} & \multicolumn2c{$\ekshell$}   &Ref.  \\ 
& & (kpc) & (km s$^{-1}$) & \multicolumn2c{(pc)} &  (km s$^{-1}$) & \multicolumn2c{($10^4$ yr)} & \colhead{($10^3~M_\odot$)} &  \multicolumn2c{($10^5$ M$_\odot$ \kms)} & \multicolumn2c{($10^{50}$ erg)} &
}
\decimals
\startdata
G34.7$-$0.4 &  W44                &  2.8  & 47       & 12.5  &  135(2)   &  2.7     & 0.39(0.01)   & 0.74(0.03)  & 1.0(0.04)  & 1 \\
G54.4$-$0.3 &                          &  6.6  &  37      & 38.4  &   62(6)    & 18       & 2.3(0.6)       & 2.0(0.6)    & 1.3(0.4)     & 1,2 \\
G69.0+2.7    &   CTB80           & 2      &  13       & 19    &  72(3)      &  7.7    & 1.1(0.2)       & 1.1(0.2)    & 0.76(0.16)   & 1 \\
G89.0+4.7    &    HB 21           & 1.7   &  $-12$  & 26    &   61(4)     & 13      &  2.2(0.4)      & 1.9(0.4)   & 1.2(0.3)  & this work \\
G189.1+3.0\tablenotemark{a}  &    IC 443          &1.5    &  $-5$    & 7.1    &  75(8)   &  2.8      &  0.49(0.06)  & 0.52(0.08) &  0.39(0.09)  &  3, 4 \\
G172.8+1.5\tablenotemark{b}   &    $\cdots$       &  1.8  &  $-20$  &  61   &  55(5)    &  33       &  5.9(0.8)      & 4.5(0.7)   &   2.5(0.6)          & 5 \\
G190.2+1.1\tablenotemark{b}   &    $\cdots$       & 8      &  20       &  88    &  77(6)    &   34      &  4.2(0.9)      & 4.5(1.0)   &   3.5(0.9)          & 6 \\
\enddata 

\tablenotetext{a}{The cap portion of \schi\ shell is not seen, so the shell expansion velocity cannot be determined from 
the \schi\ data alone. We adopt 75~\kms\ as the expansion velocity of the shell, which is the 
most prominent shock velocity obtained from the shock modeling of the 
optical line ratios in the northeastern shell \citep{alarie19}.
Most of the \schi\ mass, however, resides in the southern shell where the optical emission is faint \citep{lee08}, 
so shell expansion velocity can be smaller. }

\tablenotetext{b} {Old SNR candidates with fast expanding \schi\ shells. 
Their nature as SNRs has not been confirmed.}

\tablecomments{
$d, v_0=$ distance~and~systemic~velocity~of~SNR; 
$\rshell,\vshell=$ radius~and~expansion~velocity~of~\schi~shell;  
$\tshell(\equiv 0.3\rshell/\vshell)=$ characteristic~age~of~\schi~shell; 
$\mshell$(\schi)= \schi\ mass of shell; $\pshell,\ekshell=$ momentum~and~kinetic~energy~of~\schi~shell including the contribution from He assuming cosmic abundance.
The errors in parentheses are $1\sigma$ errors from least-squares fits  
except the error of the expansion velocity of IC 443 
which gives a dispersion in the shock speed (see Fig. 11 of \citealt{alarie19}).   
The mass, momentum, and kinetic energy are derived from the {\em thin-shell analysis}, and their uncertainties are dominated by systematic uncertainties 
in analyzing the observed data (see \S~\ref{sec:diss_error}).
Distances are the same as those used in the references except for  
G54.4$-$0.3 and HB 21. For G54.4$-$0.3, we adopt the `far' distance 
6.6 kpc corresponding to $v_0=37$~\kms\ of \citet{ranasinghe17}, which is consistent with the large 
H column density to the the source obtained from our X-ray analysis (see Appendix~\ref{sec:appendixB2}). 
For HB 21, we adopt 1.7 kpc following \cite{byun06}.
}

\tablerefs{ (1) \cite{park13}; (2) \cite{ranasinghe17}; (3) \cite{lee08}; (4) \cite{alarie19}; (5) \cite{kang12}; (6) \cite{koo06}}

\end{deluxetable}
\clearpage
 
 \newpage
\begin{deluxetable}{l c c D D l c}
\tablecaption{Parameters of Shocked Molecular Gas in Radiative SNRs with HI Shells\label{tbl-2}}
\tablewidth{0pt}
\setlength{\tabcolsep}{8pt}
\tablehead{
\colhead{Name} &  \colhead{$\mshell$(H$_2$)} & \colhead{$\vshell$} & 
\multicolumn2c{$\pshell$} & \multicolumn2c{$\ekshell$}   &Tracer  &Ref.  \\ 
 & \colhead{($10^3~M_\odot$)}  & (km s$^{-1}$) & \multicolumn2c{($10^5$ M$_\odot$ \kms)} & \multicolumn2c{($10^{50}$ erg)} & & 
}
\decimals

\startdata
W44                & 10(5)           &  12.9(0.2)          & 1.8(0.9)            &   0.23(0.12)       &HCO$^+$ $J$=1--0, CO  $J$=3--2    & 1  \\
G54.4$-$0.3\tablenotemark{a}       & ...                &  ...                     &  ...                   &   ...                & ...                                                          & ...  \\
CTB80            & ...                 &  ...                     &  ...                    &   ...               & ...                                                          & ...  \\
HB 21             & 0.73             &  $\sim 10$       &  0.16                 & 0.034                & CO $J$=1--0, 2--1                             & 2, 3 \\
IC 443            & 2                   &  25       &   0.7                & 0.17               & HCO$^+$ J=1--0                               & 4 \\
G172.8+1.5    & ...                 &  ...                    &  ...                    &   ...               & ...                                                          & ...  \\
G190.2+1.1    & ...                 &  ...                    &  ...                    &   ...               & ...                                                          & ...  \\
\enddata  

\tablenotetext{a} {There is evidence that this remnant is interacting with MC(s). See the note below.} 

\tablecomments{$\mshell$(H$_2$)= H$_2$ mass of shocked molecular gas; 
$\vshell=$ expansion~velocity~of~molecular~shell;  
$\pshell,\ekshell=$ momentum~and~kinetic~energy~of~shocked molecular gas
including the contribution from He assuming cosmic abundance. 
The errors in $\mshell$(H$_2$) and $\vshell$ of W44 are from \cite{sashida13}. 
For HB 21 and IC 443, the errors are not given in References.
}

\tablecomments{Comments on Individual Sources: {\bf W44:} \cite{sashida13} divided the kinetic energy of shocked molecular gas 
into two terms corresponding to expansion and turbulent energies. 
The quoted $\pshell$ and $\ekshell$ are those of the expansion motion. 
{\bf G54.4$-$0.3:} There are MCs along the SNR shell 
\citep{junkes92a,junkes92b,ranasinghe17}. 
\citet{junkes92a} derived the total 
mass of (1.3--2.6)$\times 10^5$~\Msun\ for these MCs and proposed that they form 
a large ($42'$ or 86 pc at 7 kpc) molecular shell expanding at 5~\kms. 
There is, however, no convincing evidence for this.
Instead, thin \h212\ emission filaments have been 
detected towards a small ($\sim 5'$), radio-enhanced area in the western 
SNR shell (\citealt{lee19}; see also Fig~\ref{fig:xray}). 
The total \h212\ luminosity ($6.4$~\Lsun) is 4\% of that of 
W44, implying a small amount of shocked H$_2$ gas.
{\bf CTB 80:} There are extended filamentary MCs superposed on the SNR, but there is no 
evidence for the interaction between the SNR and the MCs \citep{koo93}. 
{\bf HB 21:} Several small shocked MCs have been detected~\citep{koo01,byun06}. 
The listed $M({\rm H}_2)$, $\pshell$, and $\ekshell$ are the sum 
of their masses, momenta, and kinetic energies, respectively.  
The systemic LOS velocities of these MCs are in the range of a few to 20~\kms, and  the listed   
$\sim 10$~\kms\ is a characteristic expansion velocity. 
{\bf IC 443:} Both the shocked \schi\ and H$_2$ gases 
are mostly confined to an extended partial shell in the southern area (see Figure~\ref{fig:w44-ic443}). 
{\bf G172.8+1.5:} There are MCs in this area, but there is no evidence for the interaction between the SNR and the MCs 
(\citealt{kang12}; see also \citealt{dewangan18}).  
{\bf G190.2+1.1:} MCs can be seen around the SNR in the CfA CO survey \citep{dame01}, 
but there is no evidence for the interaction. 
}

\tablerefs{ (1) \cite{sashida13}; (2) \cite{koo01}; (3) \cite{byun06}; (4) \cite{dickman92}
}
\end{deluxetable}
\clearpage

 \newpage
\begin{deluxetable}{l c c c c c c}
\tablecaption{Parameters of X-ray Emitting Hot Gas in Radiative SNRs with HI Shells\label{tbl-3}}
\tablewidth{0pt}
\setlength{\tabcolsep}{8pt}
\tablehead{
\colhead{Name} &  
\colhead{$\bar n_e$} & \colhead{$T$} & \colhead{$E_{\rm th}$}   &Ref.  \\ 
& \colhead{(cm$^{-3}$)} & \colhead{(keV)} & \colhead{($10^{50}$ erg)} &
}

\startdata
W44                  & 0.36          & 0.88      & 3.7      & 1  \\
G54.4$-$0.3     &  0.07         & 0.36      & 8.3      & this work  \\
CTB 80             &  0.08         & 0.18       & 0.6    & this work \\
HB 21               &  0.03         & 0.66        & 2.4       & 2 \\
IC 443               & 0.8--2.5    & 0.29--0.64          & 1.8       & 3 \\
G172.8+1.5      &  $6\times 10^{-3}$  &  0.7                  &  5      & 4 \\
G190.2+1.1      &  $\cdots$                 &  $\cdots$          &  $\cdots$    & $\cdots$  \\
\enddata

\tablecomments{
$\bar n_e, T = $ average electron density and temperature of X-ray emitting hot gas;  
$E_{\rm th}=$ thermal energy derived by assuming that 
hot gas in pressure equilibrium is filling the spherical volume of the SNRs. 
}

\tablecomments{Comments on Individual Sources: 
{\bf W44:} Centrally-peaked diffuse X-ray emission is filling the interior of the SNR. 
The parameter $\bar n_e$ is obtained by assuming that 
the observed $EM$ ($1.8\times 10^{58}$~cm$^{-3}$, \citealt{harrus97}) is from an 
ellipsoidal volume of radii $27\times 16\times 25$ pc$^3$.
{\bf G54.4$-$0.3:} Faint diffuse X-ray emission is filling the central area of the radio SNR. 
The parameters are derived in this work using 
the archival {\it ASCA} GIS data (see Appendix \ref{sec:appendixB1}). 
{\bf CTB 80:} Faint X-ray emission of non-uniform brightness 
is filling the interior of the SNR shell. 
The parameters are derived in this work using the archival {\it ROSAT} PSPC 
data (see Appendix \ref{sec:appendixC}). 
{\bf HB 21:} Diffuse, centrally-brightened X-ray emission is filling the interior of the SNR.
The parameter $\bar n_e$ is obtained by assuming that 
the observed $EM$ ($1.5\times 10^{56}$~cm$^{-3}$, \citealt{pannuti10}) is 
from a cylindrical volume with an angular radius $11\farcm5$ and a path length $104'$. 
{\bf IC 443:} Bright X-ray emission is filling the northeastern area of the 
SNR with a faint emission extended to the southwest (Figure~\ref{fig:w44-ic443}).
The parameters $(\bar n_e,T)$ are those of the `cold' component 
of \citet{troja06}, which has been proposed to be the 
shock-heated ambient medium. They assumed that this component is 
confined to two hemispherical thin (1 pc) shells;   
eastern shell of radius 7 pc (Shell A) and a western shell of radius 11 pc (Shell B)  (see their Fig. 9). 
According to their result, the electron densities 
of Shell A and Shell B are $\sim 2$~cm$^{-3}$ and $\sim 1$~cm$^{-3}$, respectively, 
while their temperatures are both about 0.3 keV. 
We derived the thermal energy of the SNR by assuming that the inner volumes of Shell A and Shell B are   
filled with hot gas in pressure equilibrium at $\bar n_e T\sim 0.6$~cm$^{-3}$ keV and 
$\bar n_e T\sim 0.3$~cm$^{-3}$ keV, respectively. 
{\bf G172.8+1.5:} Faint extended X-ray emission is visible superposed on the SNR in the {\em ROSAT Survey Diffuse X-ray Background Map} \citep{snowden97}. The listed parameter values are crude estimates 
derived from the broad-band counts assuming that the hot gas is uniformly 
filling the interior of the SNR \citep{kang12}. 
Here we assume $n_e=1.2 n_{\rm H}$ and $E_{\rm th}\sim 3\bar n_e kT V$, and  
the numbers are sligthly different from those of \citet{kang12}.
{\bf G190.2+1.1:} The SNR is not visible in the {\em ROSAT} all sky X-ray images \citep{koo06}. 
}

\tablerefs{ (1) \cite{harrus97}; (2) \cite{pannuti10}; (3) \cite{troja06}; (4) \cite{kang12} 
}

\end{deluxetable}
\clearpage
 
 \newpage
\begin{deluxetable}{l c c c c c c c c c }
\tablecaption{Global Parameters of Radiative SNRs with HI Shells\label{tbl-4}}
\tablewidth{0pt}
\setlength{\tabcolsep}{8pt}
\tablehead{
\colhead{Name} &  \multicolumn{2}{c}{$\nbarc$} &  \multicolumn{3}{c}{$\pshell$} &\multicolumn{3}{c}{$\ekshell$} & 
\colhead{$\eth$} 
\\
& \multicolumn{2}{c}{(cm$^{-3}$)} & \multicolumn{3}{c}{($10^5$ M$_\odot$ \kms)} & \multicolumn{3}{c}{($10^{50}$ erg)} & 
($10^{50}$ erg) \\
\cline{2-3} 
\cline{4-6} 
\cline{7-9}
& \schi & \schi+H$_2$ & \schi & H$_2$ &  {\sc Hi}+H$_2$ & \schi &  H$_2$ &  {\sc Hi}+H$_2$ &  
}
\decimals
\startdata
W44                           & 1.9  &  51       &  0.74          & 1.8          &  2.54          & 1.0             & 0.23         & 1.23           & 3.8    \\
G54.4$-$0.3             &  0.4 &  0.4     & 2.0               & $\cdots$  & 2.0             & 1.3             & $\cdots$  &  1.3            & 8.3    \\
CTB80                       & 1.5  &  1.5     & 1.1              & $\cdots$  & 1.1             & 0.76           & $\cdots$  & 0.76          & 0.6   \\
HB 21                        & 1.2  &  1.6     &  1.9             & 0.1           & 2.0             & 1.2             & 0.01            & 1.21       & 2.4    \\
IC 443                       & 13    & 67       & 0.52             & 0.7           & 1.22          & 0.39           & 0.17           & 0.56       & 1.8    \\
G172.8+1.5              & 0.25  & 0.25   &    4.5            &$\cdots$   & 4.5             & 2.5             &  $\cdots$  & 2.5          & 5  \\
G190.2+1.1              & 0.06  & 0.06   &    4.5            & $\cdots$  & 4.5             & 3.5             & $\cdots$ & 3.5            & $\cdots$   \\
\enddata 

\tablecomments{$\nbarc$ = $\mshell$(\schi)$/m_{\rm H}V_{\rm sh}$ and $(\mshell$(\schi)$+M({\rm H}_2))/m_{\rm H}V_{\rm sh}$ 
where $m_{\rm H}$ is the mass of H atom and $V_{\rm sh}=4\pi \rshell^3/3$ 
(Note that these are not necessarily equal to the mean density of the ambient medium for 
SNRs in complex environment. See text.); 
$\pshell$ = momentum of the atomic ({\sc Hi}), molecular (H$_2$), and atomic+molecular ({\sc Hi}+H$_2$) SNR shell; 
$\ekshell$ = kinetic energy of the atomic ({\sc Hi}), molecular (H$_2$), and atomic+molecular ({\sc Hi}+H$_2$) SNR shell; 
$\eth$ = thermal energy of SNR from X-ray observation (see Tables~\ref{tbl-1}--\ref{tbl-3}).}

\end{deluxetable}
\clearpage

 \newpage
\begin{deluxetable}{l c c c c c}
\tablecaption{Summary of Explosion Energy \label{tbl-5}}
\tablewidth{0pt}
\setlength{\tabcolsep}{8pt}
\tablecolumns{5} \tablehead{
\colhead{Name} &  {$\nbarc$} & $\esnpsh$ & $\esnksh$ & $\esnth$ \\
 & {(cm$^{-3}$)} & \multicolumn{3}{c}{($10^{51}$ erg)}  
}
\decimals
\startdata
W44             &   1.9                     & 0.32  & 0.35  & 0.63 \\
                &   51 (\schi+H$_2$)  & 1.37  & 0.87  & 5.26 \\
G54.4-0.3       &   0.40                & 0.58  & 0.57  & 3.80 \\
CTB80           &   1.5                   & 0.35  & 0.35  & 0.47 \\
HB21            &   1.2                    & 0.61  & 0.61  & 2.01 \\
                &   1.6 (\schi+H$_2$) & 0.66  & 0.67  & 2.42 \\
IC443           &   13                      & 0.21  & 0.21  & 1.30 \\
                &   67 (\schi+H$_2$)  & 0.62  & 0.49  & 3.69 \\
G172.8+1.5      &   0.25              & 1.27  & 1.26  & 3.84 \\
G190.2+1.1      &   0.06              & 1.22  & 1.25  & \nodata 
\enddata 

\tablecomments{$\nbarc$ = number density of hydrogen nuclei 
derived from observation either \schi\ only or \schi + H$_2$ (see Table \ref{tbl-2}); 
$\esnpsh$, $\esnksh$, and $\esnth$ = SN explosion energy
to match the observed $\pshell$, $\ekshell$, and $\eth$, respectively.}

\end{deluxetable}

\clearpage
\newpage
\begin{deluxetable}{l c c c c c}
\tablecaption{Physical Parameters of the Simulated SNR \label{tbl-6}}
\tablewidth{0pt}
\setlength{\tabcolsep}{8pt}
\tablecolumns{6} \tablehead{
\colhead{Parameter} & {Entire SNR} & \multicolumn{2}{c}{Neutral} & {Ionized} & {Hot} \\
& & \colhead{$v_r <50$~\kms} & {$v_r \ge 50$~\kms} & &  
}
\decimals
\startdata       
Age (yr)            &          $5\times 10^4$        & $\cdots$    & $\cdots$  & $\cdots$  & $\cdots$ \\
$\bar n_{\rm bg}({\rm H})$ (cm$^{-3}$) & 8.5 &  $\cdots$    & $\cdots$  & $\cdots$  & $\cdots$ \\
$R_{s,{\rm vol}}$ (pc) & 16.5          & $\cdots$    & $\cdots$  & $\cdots$  & $\cdots$ \\ 
$M_{\rm H} (M_\odot)$  &    $3.81\times 10^3$            &     $3.27\times 10^3$    &    496   &     29    &      19 \\
Momentum $(10^5\ M_\odot\,{\rm km~s}^{-1})$  &  1.33   &     0.69   &    0.57   &   0.043    &    0.022 \\
Kinetic energy ($10^{49}$ erg)     &     7.78   &   1.74   &   5.24   &   0.53   &   0.27 \\
Thermal energy ($10^{49}$ erg)  &    2.03    & $\cdots$    &  $\cdots$    &  0.29    & 1.74 \\
\enddata 
\tablecomments{$\bar n_{\rm bg}({\rm H})$ = average hydrogen density of the 
background medium {\em swept-up by the SN blast wave}. Note that it is 
a little smaller than $\nbarc$ (=10 cm$^{-3}$) of the entire simulation box.
$R_{s,{\rm vol}}$ = volume-averaged radius (i.e., volume = $(4\pi/3)R_{s,{\rm vol}}^3$).
Momentum and kinetic energy include the contribution from He assuming the cosmic
abundance of $n({\rm He})/n(\rm {H})=0.1$. 
Thermal energy is obtained from $E_{\rm th} = 3\int n_e k_B T\, dV$ where $n_e=1.2n_{\rm H}$.
}
\end{deluxetable}

\clearpage
 \newpage
\begin{deluxetable}{l c c c}
\tablecaption{Global Parameters of the Simulated SNR from \schi\ 21 cm Line Analysis \label{tbl-7}}
\tablewidth{0pt}
\setlength{\tabcolsep}{8pt}
\tablecolumns{4} \tablehead{
\colhead{Parameter} & {Simulated SNR} & 
\multicolumn{2}{c}{\schi\ 21 cm line analysis} \\
& \colhead{$v_r\ge 50$~\kms} & {$|v_{\rm LOS}|\ge 70$~\kms} & {$\ge 90$~\kms}  
}
\decimals
\startdata       
Age  (yr)                          &   $5\times 10^4$       &     $5.5\times 10^4$  & $4.7\times 10^4$ \\ 
$R_{\rm sh}$ (pc)          & 16.5                           &  17.8\tablenotemark{a}      & 17.8\tablenotemark{a}    \\
H mass ($M_\odot$)      &   496                          &   513(63)                      &   330(78) \\  
$\nbarc$ (cm$^{-3}$)\tablenotemark{b}              &  1.1                          &   0.90(0.11)                   &  0.57(0.21) \\
$\bar v_{\rm exp}$         &   82                            &   94(6)                           &   112(6) \\
Momentum $(10^4\ M_\odot~{\rm km}\,{\rm s}^{-1})$   &    5.73         &  6.8(0.9)         &  5.1(1.3) \\   
Kinetic Energy $(10^{49}\ {\rm erg})$  &  5.24  &  6.3(1.1)   & 5.7(1.5) \\
%
\enddata 

\tablenotetext{a} {Geometrical mean radius of the simulated SNR projected on the sky 
(i.e., projected area = $\pi R_{\rm sh}^2$). We assume that this radius 
has been obtained from other observations, e.g., from a radio continuum observation, and 
use it to obtain the age in \schi\ 21 cm line analysis (Equation \ref{eq:tsh}).
}

\tablenotetext{b} {$\nbarc$ = H mass$/(4\pi R_{\rm sh}^3/3)$}

\tablecomments{The global parameters derived from an \schi\ analysis have been compared 
with those of the fast ($v_r\ge 50$~\kms) neutral component of the simulated SNR.
The two columns in the \schi\ 21 cm line analysis show the parameters obtained from 
a fit to the profiles at $|v_{\rm LOS}|\ge 70$~\kms\ and $|v_{\rm LOS}|\ge 90$~\kms\ (see Figure~\ref{fig:ssnr_hi}), respectively. 
The velocity $\bar v_{\rm exp}$ of the simulated SNR is the 
mass-weighted average radial velocity.
Momentum and kinetic energy include the contribution from He assuming the cosmic
abundance of $n({\rm He})/n(\rm {H})=0.1$. The errors in parenthesis are formal errors from the fit.
}
\end{deluxetable}

\clearpage
\newpage
\begin{deluxetable}{l c c c c }
\tablecaption{Explosion Energy of the Simulated SNR from \schi\ 
21 cm Line Analysis\label{tbl-8}}
\tablewidth{0pt}
\setlength{\tabcolsep}{8pt}
\tablecolumns{5} \tablehead{
\colhead{\schi\ 21 cm Line Analysis} &  {$\nbarc$} & $\esn(\pshell)$ & $\esn(\ekshell)$ & $\esn(\eth)$ \\
 & {(cm$^{-3}$)} & \multicolumn{3}{c}{($10^{51}$ erg)}  
}
\decimals
\startdata
$|v_{\rm LOS}|\ge 70$~\kms   &   0.90     &  0.23     &  0.23    &  0.17  \\  
$|v_{\rm LOS}|\ge 70$~\kms  + Slow component                &   6.7      &   0.54    &  0.47    &  0.63 \\
\enddata 

\tablecomments{The first row shows the results when we use the parameters of the \schi\ shell  
derived from the analysis of the \schi\ profile at $|v_{\rm LOS}|\ge 70$~\kms\ in Table \ref{tbl-7}.
The second row show the results when the 
slow component ($|v_r<50$~\kms) has been detected, e.g., in molecular lines, 
and their momentum and kinetic energy in Table~\ref{tbl-6} 
are included in addition to those of the \schi\ shell.  (See text for more details.)
}

\end{deluxetable}

\clearpage
\newpage
\appendix

\section{Physical Parameters of the Expanding \schi\ Shell in HB 21}\label{sec:appendixA}

\begin{figure}[ht]
\epsscale{1.0}
\centering
\vspace{0.5truecm}
\plottwo{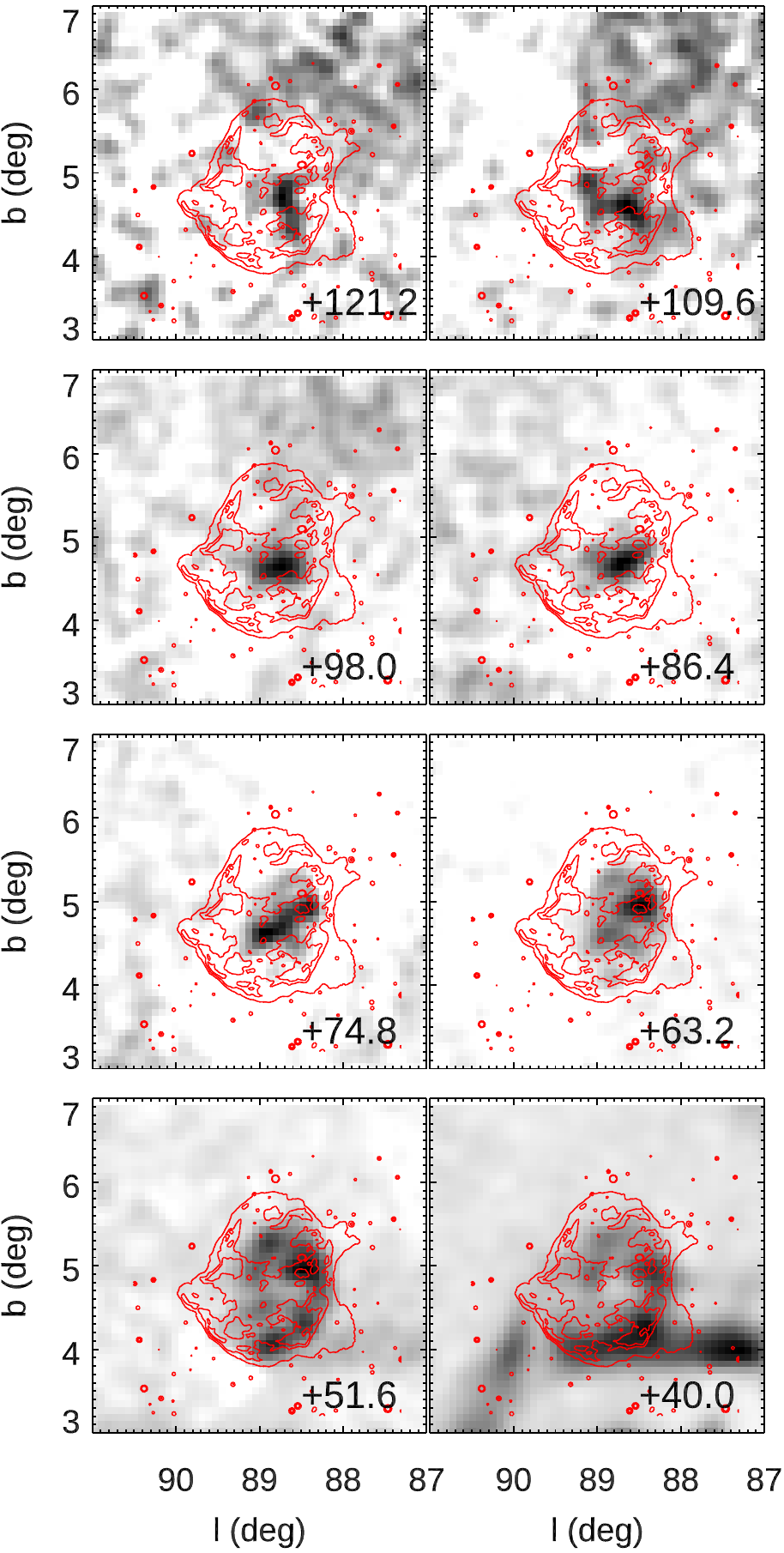}{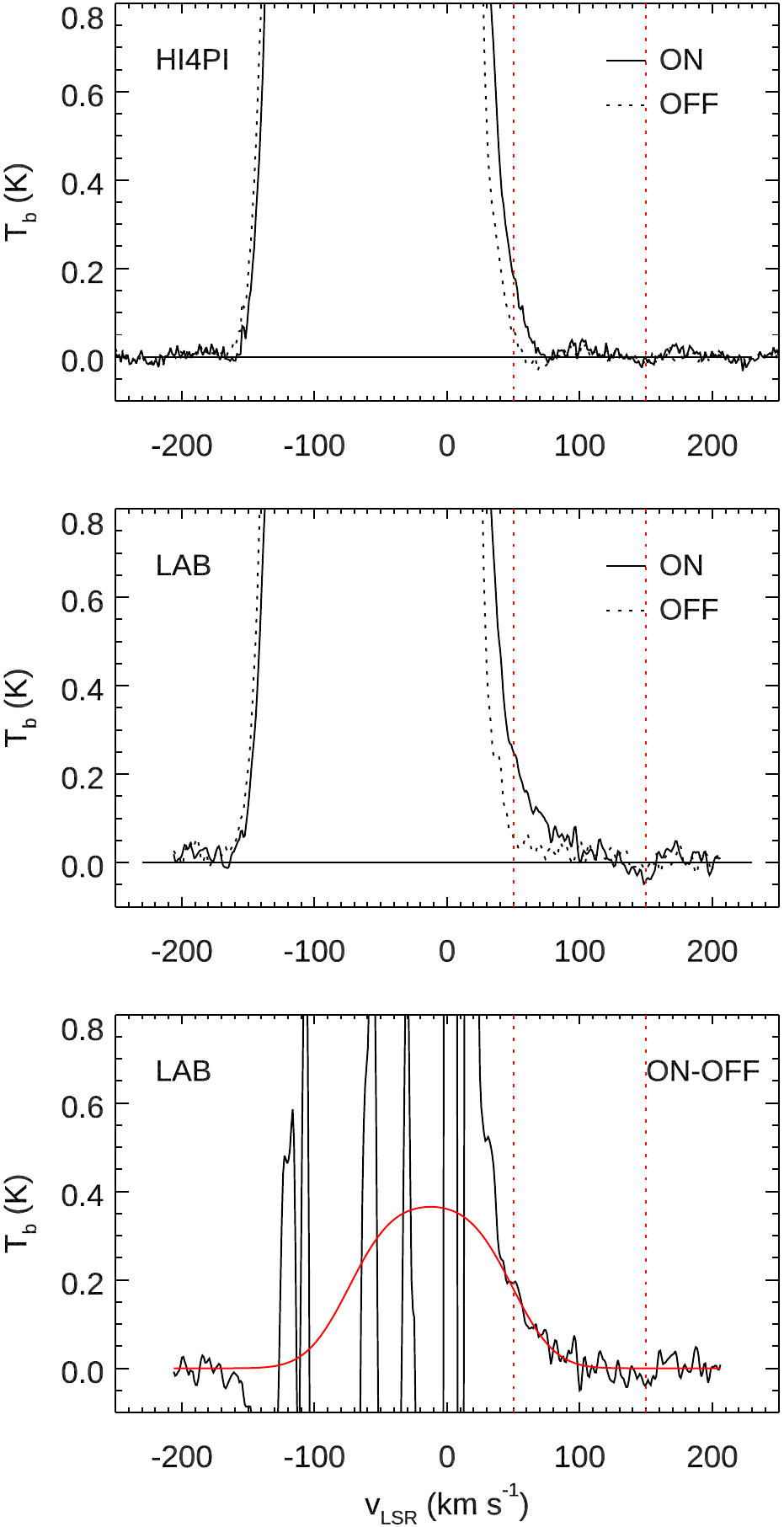}
\caption{{\bf Left:} \schi\ channel maps of the fast-expanding \schi\ shell in HB 21. 
Red contours show the SNR morphology in radio continuum obtained 
by the Dominion Radio Astrophysical Observatory (DRAO) \citep[see][]{byun06}.
Each map has been obtained by integrating over $11.6$~\kms\ 
centered at the labeled $\vlsr$ (\kms) and its 
gray scale is normalized by the maximum brightness.  
{\bf Top right:} Source and background \schi\ 21 cm profiles of HB 21 from the HI4PI data. 
The source profile is the average profile inside the 
circular area of radius $1\fdg1$ centered at $(l,b)=(88\fdg92,4\fdg75)$, while   
the background profile is the average profile inside the annulus of 
inner and outer radii of $1\fdg4$ and $1\fdg7$, respectively.  
{\bf Middle right:} Same as the top right but from the LAB data. 
{\bf Bottom right:} Background-subtracted \schi\ spectrum of HB 21 from the 
LAB data and a fit. Variations in \schi\ 21 cm brightnesses between the source and background
directions cause the wild fluctuations in this background-subtracted profile at low velocities. 
The red solid line is a best fit to the profile between the velocity range 
marked by the red dotted lines (i.e., 50--150~\kms).   
}
\label{fig:HB21}
\end{figure}

In this Appendix, we derive the parameters of the    
\schi\ shell associated with the SNR HB 21 (G89.0+4.7). 
HB 21 is an old SNR with an irregular
radio shell of large angular size ($120'\times 90'$; see Figure~\ref{fig:HB21}).
An expanding \schi\ shell associated with the SNR was detected by 
\cite{koo91} using the Hat Creek 85 foot telescope (FWHM=$36'$).
They detected high-velocity gas moving at $\vlsr=40$ to 120~\kms\ 
in the southern part of the SNR and derived the parameters of an expanding shell 
matching the observed \schi\ 21 cm line emission properties.
We use the data from the HI4PI \citep{hi4pi16} and  
the Leiden/Argentine/Bonn (LAB; \citealt{kalberla05} \schi\ 21 cm line all-sky surveys. 
The HI4PI survey, which has been constructed from  
the Galactic All-Sky Survey (GASS) and the Effelsberg-Bonn HI Survey (EBHIS),
provides a higher resolution ($16\farcm2$) map with a comparable sensitivity ($1\sigma\sim 43$ mK). 
But as we will see below, the spectra appear to have baseline fluctuations at low-intensity level, 
so that we use the LAB survey data for quantitative analysis. 
The LAB survey combined three independent surveys of angular resolution of 30$'$--36$'$
to produce all-sky data at $0\fdg5$ pixels with an rms noise of 0.07--0.09~K 
\citep{kalberla05,hartmann97,bajaja05,arnal00}.
The distance to HB 21 has been estimated to be 
1.7 kpc and its systemic velocity to be $-12$~\kms\ 
\citep{byun06}.

Figure~\ref{fig:HB21} shows channel maps from $\vlsr=+40$ to +121~\kms\ 
where we can clearly see \schi\ emission confined inside the SNR boundary.
At the highest positive velocities (+121 to +75~\kms), a centrally-peaked compact  
emission feature is seen in the central area. 
At lower velocities, the emission becomes more extended and forms a ring-like shape. 
The ring morphology becomes prominent at $\vlsr=52$~\kms. 
The ring-like emission feature persists at lower velocities ($\vlsr\simlt 40$~\kms), 
but there are other emission features   
superposed on the southern part of the SNR. 
Note that the SNR HB 21 is at $l=89\degr$ where the 
LSR velocity due to the Galactic rotation is mostly negative. 
Therefore, the \schi\ emission at large $\vlsr(\gg 0)$  
in Figure~\ref{fig:HB21} indicates local energy injection, and the positional coincidence and the velocity structure 
strongly suggest that we are observing a receding portion of 
an fast expanding shell associated with the SNR.

In the right frame of Figure~\ref{fig:HB21}, we show the average \schi\  profile  
toward the SNR together with a background spectrum. 
The excess emission above the background emission 
at $\vlsr\ge 30$~\kms\ is clearly seen. 
The amount of the excess emission, however, is different between 
the HI4PI and the LAB profiles.
At $\vlsr=50$--100~\kms, for example, the excess emission 
in the HI4PI profile is considerably weaker than 
that in the LAB profile. 
We note that the HI4PI spectra of the area outside the SNR  
show negative dips in this velocity range, so that  
the difference is most likely due to the background fluctuation in the 
HI4PI spectra. 
We therefore use the LAB spectra for quantitative analysis.
As we mentioned above, at $\vlsr<0$, the emission from \schi\ gas 
in the Galactic disk along the LOS is strong,   
and the approaching portion of the shell cannot be confirmed.

We derive the shell parameters from the background-subtracted LAB profile 
in Figure~\ref{fig:HB21}. We assume that the shell is expanding at 
a constant systemic speed $\vshell$ but has a considerable 
velocity dispersion (FWHM=50~\kms). 
Previous high-resolution \schi\ observations showed that 
the expanding shells in SNRs are clumpy and the emission lines of 
resolved clumps are broad ($\sim 50$~\kms),  
presumably due to inhomogeneity of the ambient medium and/or  
turbulent motions produced by hydrodynamic instabilities \citep[][and references therein]{park13}. 
With the systemic velocity fixed at $v_0 = -12$~\kms, then 
we can fit the observed \schi\ profile as the emission from an expanding shell. 
The details may be found in \cite{park13}.
The fit was done using the IDL routine MPFITFUN \citep{markwardt09}.
The best fit profile is shown in Figure~\ref{fig:HB21}. 
Its parameters are $\vshell=61.4\pm 3.5$~\kms\ and $T_{b, {\rm max}}=0.37\pm 0.06$~K.
The average column density of the expanding shell 
obtained from the fit is   
$\bar N_{\rm H}=1.822\times 10^{18}~T_{b, {\rm max}}\times 2\vexp=4.14\pm0.71\times 10^{19}$~cm$^{-2}$ 
assuming that the \schi\ emission is optically thin.
Note that this includes both receding and (unobservable) approaching sides of the shell. 
The basic idea is that the momentum and kinetic energy imparted to the approaching side might be comparable to those of the receding side. 
(See \S~\ref{sec:diss_error} for a discussion about the uncertainty in this thin-shell analysis.)
If we multiply $\bar N_{\rm H}$ by the area 
($\pi\times(1\fdg1)^2$) used to extract the average spectrum,  
the \schi\ mass of the shell becomes $\mshell=2.22 \pm 0.39\times 10^3$~\Msun. 
And the momentum and kinetic energy of the shell (including the contribution from He) become 
$\pshell=\mshell\vshell=1.91 \pm 0.35 \times 10^{5}$~\Msun~\kms\ and 
$\ekshell=\mshell\vshell^2/2=1.17 \pm 0.25 \times 10^{50}$~erg.
(For comparison, \cite{koo91}  obtained $\mshell=1.9\times 10^3$~\Msun\ and 
$\ekshell=2.9\times 10^{50}$~erg. They used $v_0=-1$~\kms\ and 
the maximum velocity 124 \kms\ as the expansion speed of the shell.) 
The resulting dynamical age of the shell is $\tshell=0.3 R_s/v_s\approx 1.3\times 10^5$~yr
where we used a geometrical mean radius (26 pc)
as the radius of the SNR ($R_s$). 

\newpage
\clearpage

\section{Thermal Energy and Distance of G54.4$-0.3$}\label{sec:appendixB}

\subsection{Thermal Energy}\label{sec:appendixB1}

X-ray emission from SNR G54.4$-$0.3 has been detected with {\it ROSAT}, {\it ASCA}, 
and {\it Chandra}. G54.4$-$0.3 shows diffuse X-ray emission with a centrally-peaked 
morphology surrounded by a roughly circular radio shell with a radius of $R$ $\sim$ 
20$'$ (Figure~\ref{fig:xray}a, Park et al. 2013). Taking advantage of the large 
field of view (50$'$ in diameter) and the broad spectral coverage ($E$ $\sim$ 0.3 -- 
10 keV), we use the archival {\it ASCA} Gas Imaging Spectrometer (GIS) data to 
estimate the thermal energy of 
G54.4$-$0.3. Three ObsIDs in the {\it ASCA} pulic archive cover the entire G54.4$-$0.3 
(Figure~\ref{fig:xray}a). We extracted the source spectrum from a circular region with 
an angular radius $R$ = 12$'$, centered on RA(J2000) = 19$^{\rm h}$ 33$^{\rm m}$ 10$^{\rm s}$.92, 
Dec(J2000) = 19$\degr$ 03$'$ 15$\farcs$7 (the yellow circle in 
Figure~\ref{fig:xray}a) in ObsID 52000000 (pointing at the central 
X-ray emission feature of G54.4$-$0.3, the middle cyan circle in Figure~\ref{fig:xray}a). 
This region includes the bulk of the centrally-peaked X-ray emission in this SNR. 
The soft X-ray image from {\it ROSAT} data shows a pointlike emission feature near 
the center of the SNR (Figure~\ref{fig:xray}a). This source is faint in the {\it ASCA} 
GIS data, whose fractional contribution in our source spectrum is negligible ($\sim$5\% 
in the 0.5 -- 9 keV band flux, assuming a source region with a radius of 2$^{\prime}$). 
Thus, we did not remove this source in our source spectral extraction. We extracted the 
background spectrum from a circular region just outside of the northern radio shell 
based on ObsID 52001000 (the white circle in Figure~\ref{fig:xray}a). For our spectral 
extraction, we used the standard pipeline-processed event files in the {\it ASCA} public 
archive. We combined GIS2 and GIS3 data for both of the source and background spectra.
The total effective exposures are $\sim$37 ks and $\sim$34 ks for the extracted source 
and background spectra, respectively. Based on these data, we collected $\sim$4300 
background-subtracted source counts in the 0.6 -- 8 keV band.

The extracted X-ray spectrum of G54.4$-$0.3 can be fitted with two-component spectral 
models for thermal or non-thermal origins. For the thermal
component, we consider X-ray emission from the shocked optically-thin hot gas in either
collisional ionization equilibrium (CIE, the APEC model in XSPEC [\citealt{arnaud96}]), or
non-equilibrium ionization (NEI, the ``plane-shock'' model, PSHOCK [\citealt{borkowski01}] 
in XSPEC). We fit the non-thermal component with a simple power law (PL) model. 
For our spectral model fits, we re-grouped the X-ray spectrum of G54.4$-$0.3 to achieve 
a minimum of 20 counts per photon energy channel. The observed X-ray spectrum of G54.4$-$0.3
can be fitted with either APEC + PSHOCK or APEC + PL models. In these spectral
model fits, the soft thermal APEC component may be replaced with a PSHOCK component.
In such a case, because the best-fit soft PSHOCK component model indicated large values 
of the ionization timescale of $\tau > 10^{13}$ cm$^{-3}$ s ($\tau$ = $n_et$, where 
$n_e$ is the postshock electron density and $t$ is the time since the gas has been 
shocked), which is equivalent with a CIE condition, we consider the APEC model for 
the soft component X-ray emission in the following discussion. Both of the best-fit 
APEC + PSHOCK and APEC + PL models are statistically acceptable with equally good 
fits ($\chi^2/\nu = 206.1/194$ for APEC + PSHOCK, and $\chi^2/\nu = 213.0/195$
for APEC + PL). We find that the best-fit elemental abundance is generally low ($\la$1 
solar with respect to the solar abundance by \citet{anders89}). Thus, we 
fixed the abundance at the solar value in our spectral model fits, which is in line 
with the low-abundant interstellar origin for the observed X-ray emission in this 
old SNR. The best-fit parameters for our APEC + PSHOCK model fit are: the absorbing 
foreground H column $N({\rm H})_{\rm X-ray} = 1.07^{+0.13}_{-0.16} \times 10^{22}$ cm$^{-2}$, 
the electron temperature $k_B T_{\rm APEC} = 0.33^{+0.09}_{-0.06}$ keV,  
$k_B T_{\rm PSHOCK} = 10.5^{+5.8}_{-3.0}$ keV, 
the ionization timescale $\tau = 2.3^{+2.5}_{-2.1} \times 10^{9}$ cm$^{-3}$ s. 
Our best-fit APEC + PL model shows similar results for the soft APEC component 
($k_B T_{\rm APEC} = 0.38^{+0.12}_{-0.08}$ keV absorbed by 
$N({\rm H})_{\rm X-ray} = 1.01\pm 0.15 \times 10^{22}$ cm$^{-2}$). 
The best-fit PL component indicates the photon
index $\Gamma = 1.53^{+0.23}_{-0.26}$. In Figure~\ref{fig:xray}b we show the 
{\it ASCA} GIS spectrum of G54.4$-$0.3 with the best-fit APEC+PSHOCK model overlaid.

The spectrally-soft component from the hot gas in CIE 
condition (fitted with $k_B T \sim 0.3$ keV) may be reasonable for X-ray emission 
from an old SNR. However, it is difficult to interpret the hard component emission 
with an extremely high electron temperature of $k_B T \sim 10$ keV with a very low 
$\tau$ as suggested by our APEC + PSHOCK model fit. 
The estimated electron temperature is similar to that of the 
Galactic ridge X-ray emission (GRXE), i.e., the diffuse X-ray emission from the thin disk 
surrounding the Galactic midplane \citep{warwick85, valinia98}. 
So the GRXE may partially contribute to the hard component emission, 
although it is relatively weak at $\ell\sim 54^\circ$.
This spectrally-hard component 
of the X-ray emission is equally fitted with a PL model with $\Gamma \sim 1.5$, 
which may also allow us alternative interpretations such as the shock-accelerated 
synchrotron emission from a separate young SNR projected along the LOS. 
In fact, G54.4$-$0.3 is projected against a complex of numerous radio shells, suggesting 
its location in a region with a strong star-forming activity. The origin and physical 
interpretations of the spectrally-hard component remain puzzling. Further studies 
based on deeper X-ray observations and hydrodynamic calculations may be required 
to reveal the true nature of the hard X-ray emission in the G54.4$-$0.3 region, 
which is beyond the scope of this work. For the purposes of this work, we conclude 
that the X-ray spectrum of G54.4$-$0.3 is most likely dominated by the soft thermal 
component.

Based on the best-fit volume emission measure of the soft APEC model component 
($EM = \int n_e n_{\rm H} dV =$ (1.3--1.8)$\times 10^{58}$ cm$^{-5}$ 
where $n_{\rm H}$ is the H density), we 
estimate an average electron density $\bar n_e$. In this calculation we assumed 
$n_e \approx 1.2 n_{\rm H}$ for a fully-ionized hot gas of cosmic abundance. 
We also assume a cylindrical volume for the derived volume emission measure 
with an angular radius $12'$ and a path length
(along the LOS) corresponding to the physical diameter
of the spherical SNR ($2R = 40'$).
We adopt 6.6 kpc as the distance to the SNR (see Appendix~\ref{sec:appendixB2}).
Our estimated average electron density is $\bar n_e  \sim 0.07$ cm$^{-3}$. 
The mass of X-ray emitting gas inferred from the observed $EM$ is  
$M_X \approx 1.4m_{\rm H}\times EM/\bar n_e\sim 260$~\Msun\ where $m_{\rm H}$ is 
the mass of hydrogen nucleon.
Assuming that the interior of the SNR is filled with hot gas in pressure equilibrium, 
we estimate the thermal energy of the SNR to be 
$E_{\rm th} \approx 3 \bar n_e k_B T V_s  \sim 8 \times 10^{50} $ erg where 
$V_s$ is a spherical volume of radius $20'$ (=38 pc at 6.6 kpc).

\subsection{Distance}\label{sec:appendixB2}

\begin{figure}
\centering
\includegraphics[scale=0.6]{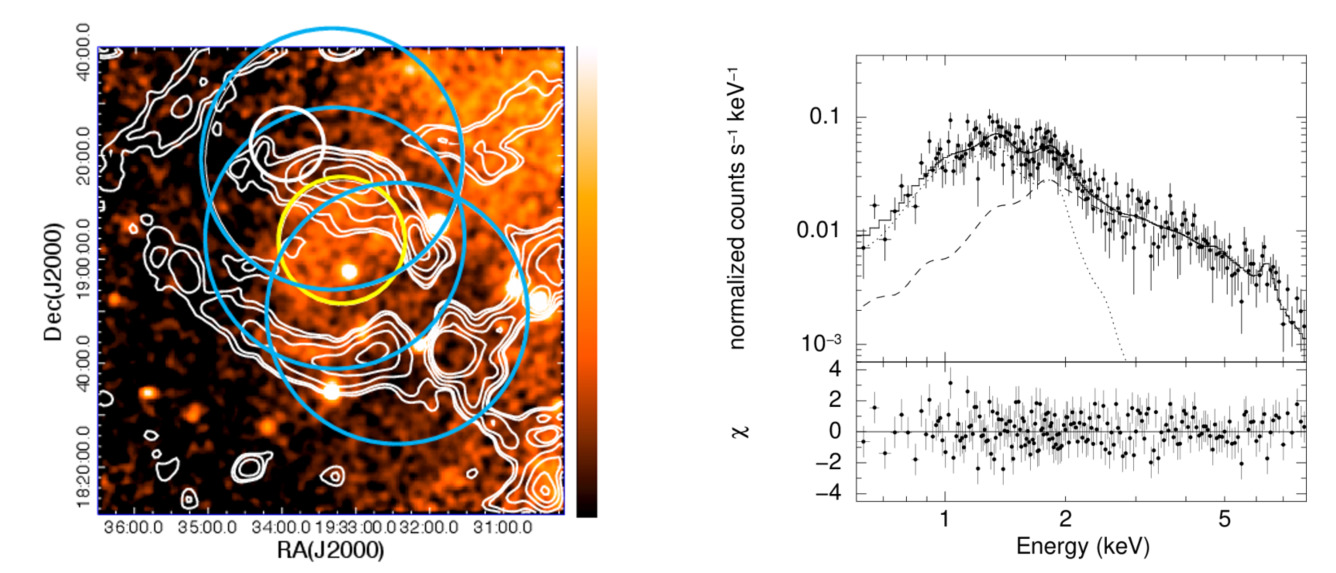}
\caption{{\bf Left:} A smoothed false-color {\it ROSAT} PSPC image of G54.4$-$0.3.
The image has been taken from NASA's {\it SkyView} on-line facility. The radio 
contours (taken from the Green Bank 4850 MHz survey) are overlaid. Cyan circles 
show the field of views of three archival {\it ASCA} GIS ObsIDs that detected
G54.4$-$0.3. Our source and background regions for G54.4$-$0.3 are marked with 
a yellow and a white circle, respectively.
{\bf  Right:} The {\it ASCA} GIS spectrum of G54.4$-$0.3. The best-fit APEC (dotted) + 
PSHOCK (dashed) model
is overlaid. The bottom panel is residuals from the best-fit model.
}
\label{fig:xray}
\end{figure}

\begin{figure}
\epsscale{0.65}
\plotone{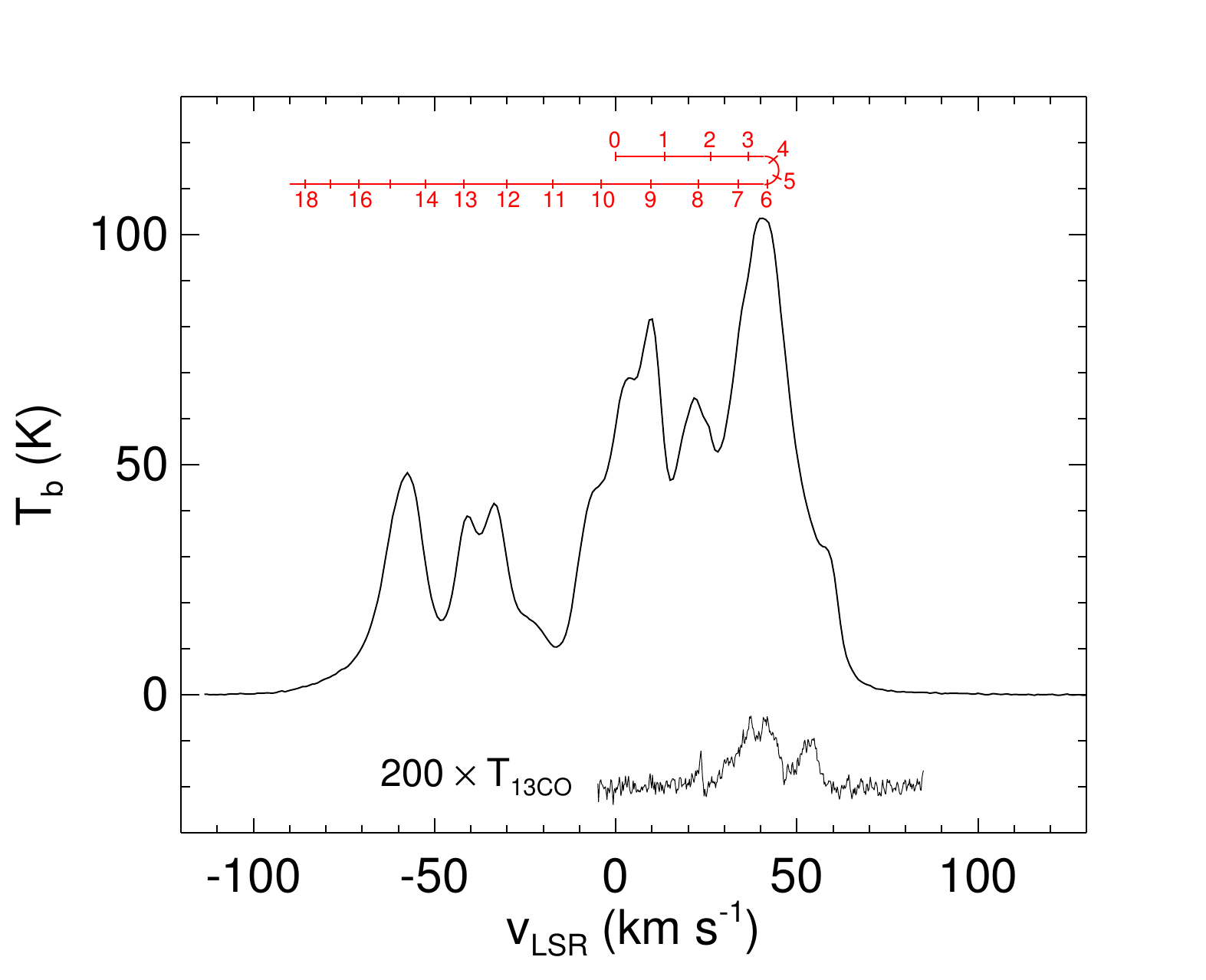}
 \caption{
Average \schi\ 21 cm and $^{13}$CO $J$=1--0 emission line profiles toward G54.4$-$0.3. The ruler on top of the profile shows how the heliocentric distance varies as a function of LSR velocity. We have adopted the Galactic rotation curve of \cite{reid14} where the Galactocentric radius of the Sun $R_0=8.34$~kpc and the Galactic rotation speed at the solar circle $\Theta_0=220$~\kms. 
}
\label{fig:hiprofile}
\end{figure}

The distance to G54.4$-$0.3 is uncertain. 
There is a CO shell-like structure surrounding the SNR at $\vlsr\sim40$~\kms, 
and the distances between the near (3.3 kpc) and far (6.6 kpc) kinematic 
distances corresponding to this velocity have been proposed \citep{junkes92a,zychova16,ranasinghe17}.
Another distance estimate can be obtained by comparing 
the X-ray absorbing column density 
$N({\rm H})_{\rm X-ray}(\sim 1 \times 10^{22}$ cm$^{-2}$; Appendix~\ref{sec:appendixB1}) 
to the \schi$+$H$_2$ column densities along the LOS to the SNR. 
In Figure~\ref{fig:hiprofile}, we show 
\schi\ 21 cm and $^{13}$CO $J$=1--0 emission line profiles toward G54.4$-$0.3 
extracted from the VLA Galactic Plane Survey \citep{stil06} and the 
Boston University-Five College Radio Astronomy Observatory Galactic Ring Survey \citep{jackson06}, respectively. 
The emission profiles are the average profiles of the X-ray emitting SNR area 
(i.e., the area bounded by the yellow circle in Fig.~\ref{fig:xray}). 
In Figure~\ref{fig:hiprofile}, the emission at $\vlsr>0$~\kms\ is basically  
from the \schi\ gas inside the solar circle, 
so that if we add up all the emission at $\vlsr>0$~\kms\, it yields the total gas column density 
up to the opposite side of the solar circle at $\sim 10$ kpc in $\ell=54\fdg4$. 
We derive $N({\rm HI})=0.92\times 10^{22}$~cm$^{-2}$ assuming \schi\ spin temperature of 150 K.
For the H$_2$ column density, we derive 
$N({\rm H}_2)\sim 4\times 10^{20}$~cm$^{-2}$ 
from the average integrated $^{13}$CO $J$=1--0 intensity of 1.3 K~\kms\  
assuming that $^{13}$CO gas is in local thermodynamic equilibrium 
(LTE) with an excitation temperature of 
10 K and $N({\rm H}_2)/N({^{13}}{\rm CO})/=5\times 10^5$
\citep[e.g.,][]{dickman78}.
This LTE $^{13}$CO column density may underestimate the true 
$^{13}$CO column density by a factor of 2--3 
\citep{padoan00,szucs16}, in which case 
the column density of H nuclei in molecular gas 
could be as much as $\sim 30$\% of the atomic gas. 
Hence, the total H column density up to 10 kpc 
in $\ell=54\fdg4$ is estimated to be about $1.2\times 10^{22}$~cm$^{-2}$, 
and, if we naively assume that about half of this column resides 
on the near side of the tangent point, 
the foreground X-ray absorbing column density 
$N({\rm H})_{\rm X-ray} \sim 1 \times 10^{22}$ cm$^{-2}$ implies that 
the SNR G54.4$-$0.3 is probably beyond the tangential point.  
This is consistent with the conclusion of \citet{ranasinghe17}  
based on their \schi\ 21 cm absorption line study.
So we adopt the far kinematic distance of the CO shell (6.6 kpc), 
as the distance of G54.4$-$0.3.

\clearpage

\section{Thermal Energy of CTB 80}\label{sec:appendixC}

\begin{figure}
\centering
\includegraphics[scale=0.55]{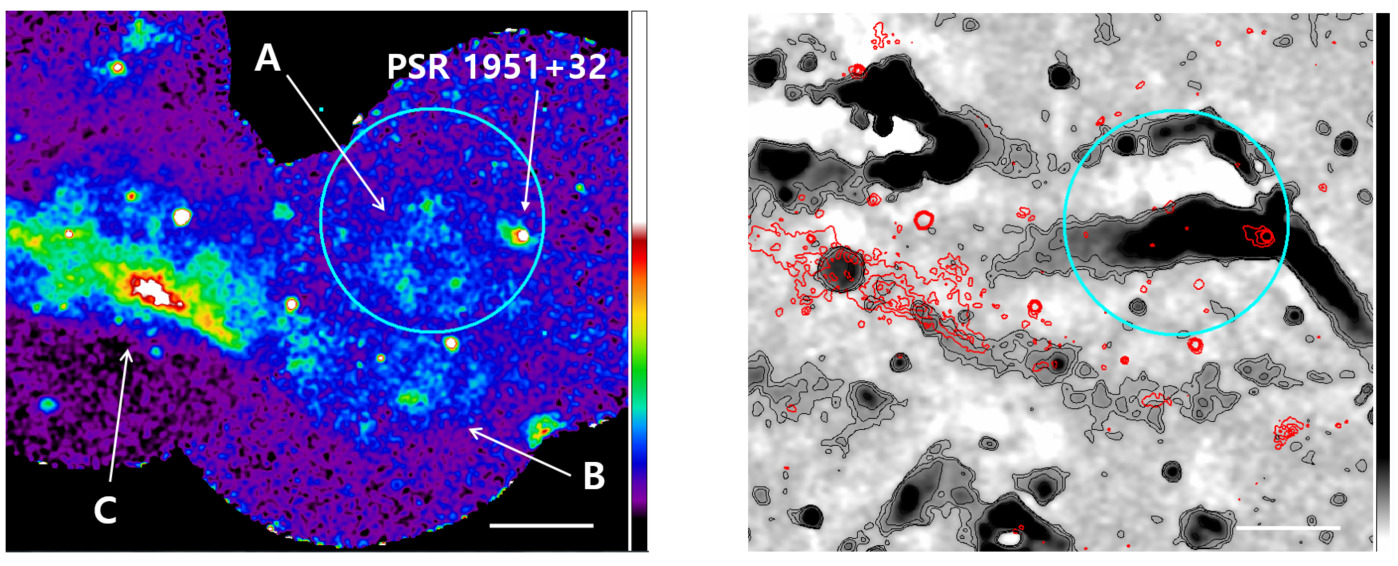}
\caption{{\bf Left:} A smoothed false-color {\it ROSAT} PSPC image of 
the $3\degr\times2\fdg6$ area surrounding CTB 80.
North is up and east is to the left. 
The image has been taken from NASA's {\it SkyView} on-line facility.
Cyan circle shows the boundary of 
the SNR defined by the IR shell \citep{fesen88,koo90}.
The X-ray features discussed in the text are marked. 
{\bf Right:} Radio 4.85 GHz image of the same area 
from the Green Bank 4850 MHz survey.
The red contours shows the X-ray brightness distribution in the left image.
In both images, the color (gray) scale is linear and the scale bar represents $0\fdg5$.
}
\label{fig:ctb80a}
\end{figure}

CTB 80 is a peculiar SNR   
with a pulsar wind nebula (PWN) embedded in an extended structure of complex morphology in radio 
(Figure~\ref{fig:ctb80a}).  
There is a 39.5 ms pulsar with the spin-down age of $1.1\times 10^5$~yr 
at the center of the PWN, suggesting that CTB 80 is an old SNR \citep{kulkarni88,fruchter88}.
This pulsar, PSR 1951+32, is located at the eastern inner boundary of  
a nearly complete spherical   
shell of $\sim 1\degr$ diameter with prominent infrared (IR) dust emission \citep{fesen88}.
The detection of fast-expanding \schi\ shell with a dynamical age of  
$\sim 8\times 10^4$~yr (at 2 kpc) matching the shape of the IR shell 
supports that the IR shell is the SNR shell of CTB 80 \citep{koo90}.
X-ray emission associated with PSR 1951+32 and its nebula has been 
studied in detail, but the study on the thermal 
X-ray emission from the SNR is scarce. 
\cite{safi-harb95} studied the X-ray emission of CTB 80 with the Position Sensitive 
Proportional Counter (PSPC) aboard {\it ROSAT}, and 
proposed that the soft X-ray emission in the southeastern area of the field 
is associated with the SNR (see Fig. \ref{fig:ctb80a}).
The X-ray emission, however, 
extends much further away beyond the SNR shell, 
and it is difficult to reconcile the spatial distribution of the X-ray emission  
with the spherical SNR shell.  
\cite{mavromatakis01} detected optical filaments 
overlapping with the X-ray emission outside the SNR shell 
and suggested that they do not appear to be related to CTB 80, while 
\cite{leahy12} studied \schi\ emission surrounding CTB 80 
and proposed that CTB 80 is a larger ($R\sim 76'$) 
SNR including the X-ray emission beyond the IR SNR shell. 

Figure \ref{fig:ctb80a} shows the 
{\it ROSAT} X-ray and the Green Bank 
4.85 GHz maps of the $3\degr\times2\fdg6$ area surrounding CTB 80. 
The cyan circle represents the location 
of the IR SNR shell of radius $32'$.
The diffuse X-ray emission features mentioned above are   
labeled by ``A'' and ``B''; source A for the X-ray emission inside the IR shell 
and source B for that outside the IR shell.  
We also see a bright extended 
X-ray filament (hereafter source ``C'').
\cite{yoshita00} showed that the X-ray emission of source C is thermal ($T\sim 0.4$~keV) 
with {\it ASCA} and {\it ROSAT}.
They noted that there is a large radio filament 
overlapping with source C and concluded that 
these radio and X-ray filaments are portions of an old SNR (see Fig. \ref{fig:ctb80a}:Right). 
We note that the radio filament extends further out to west 
and overlaps with source B. 
Hence, it seems reasonable to consider that 
sources B and C and the associated radio filament are 
the structures belong to another evolved SNR  
and that only source A is associated with CTB 80.
In the following, we derive the thermal energy of source A.

\begin{figure}
\centering
\hspace*{-1.5cm}
\includegraphics[scale=0.65]{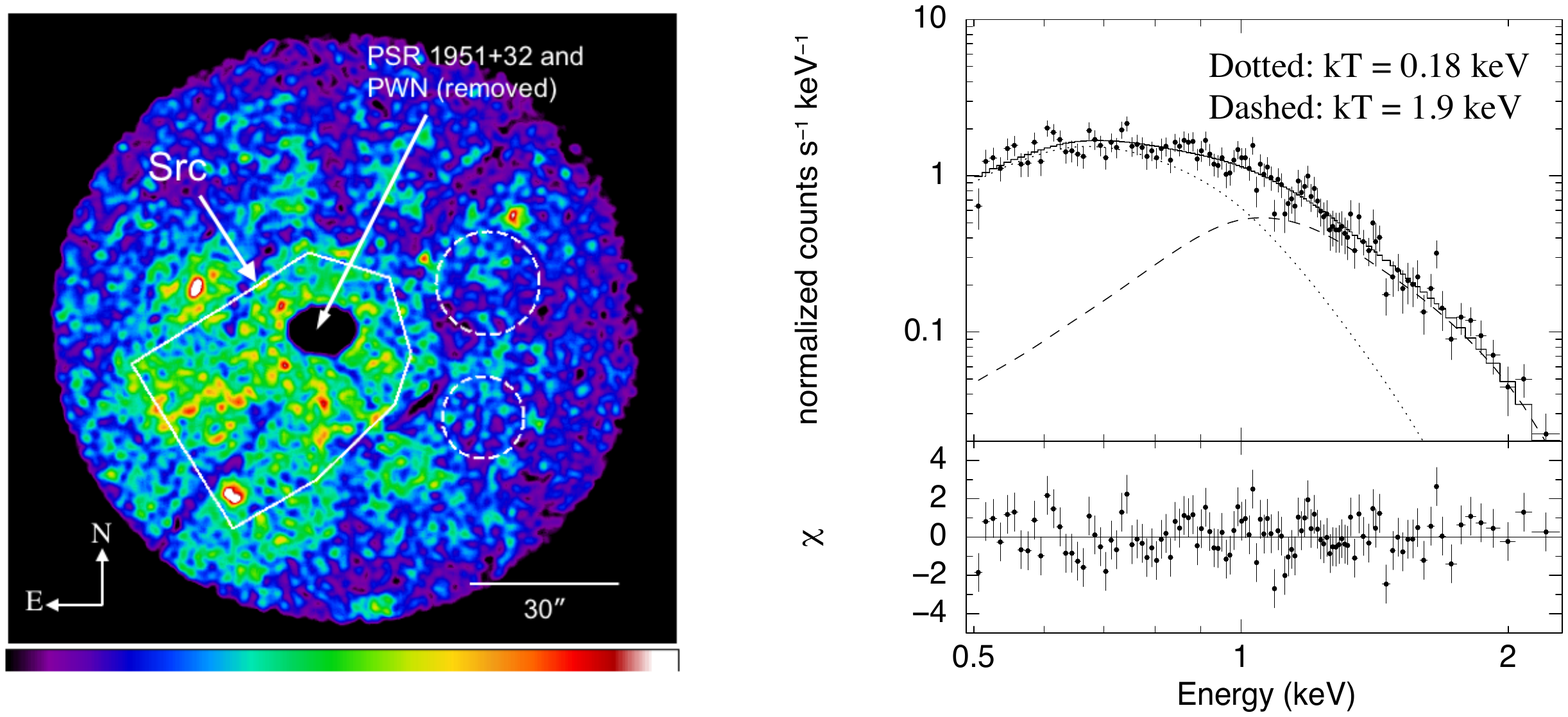}
\vspace*{-3.0cm}
\caption{{\bf Left:} A smoothed false-color {\it ROSAT} PSPC image of CTB 80.
The areas used for the spectral analysis are marked. PSR 1951+32 and its
associated PWN \citep{safi-harb95} has been removed.
{\bf  Right:} The 0.5 -- 2 keV band background-subtracted PSPC spectrum of 
CTB 80, as extracted from the Src region in the left image. The best-fit 
two-component APEC model is overlaid. The bottom panel is residuals from 
the best-fit model.
}
\label{fig:ctb80b}
\end{figure}

Based on the archival {\it ROSAT} PSPC data (ObsID RP400048A00, $\sim$5 ks 
exposure time), we extract
the X-ray spectrum of CTB 80 from a $\sim$48$'$ $\times$ 36$'$ region (the 
``Src'' region, excluding the pulsar/PWN, in Figure~\ref{fig:ctb80b}:Left) 
corresponding to the central area of source A. The background spectrum was 
extracted from ``source-free'' regions off-axis to west (dashed circles in  
Figure~\ref{fig:ctb80b}:Left).
The X-ray spectrum of CTB 80 shows significant fluxes at $E$ $<$ 0.5 keV,
for which a low foreground column ($N_{\rm H}$ $\sim$ 10$^{20}$ cm$^{-2}$) is
implied due to the short mean free path for X-ray photons at these energies. 
We note that the foreground column for PSR 1951+32, which is presumed to be associated
with CTB 80, was estimated to be $N_{\rm H} = 3\times 10^{21}$~cm$^{-2}$ 
\citep{safi-harb95}. Indeed we find that the {\it ROSAT} PSPC spectrum of 
PSR 1951+32 shows negligible fluxes at $E$ $<$ 0.5 keV.
Assuming the same $N_{\rm H}$ = 3 $\times$ 10$^{21}$ cm$^{-2}$ for CTB 80, 
we conclude that the observed X-ray flux 
at $E$ $<$ 0.5 keV should be the foreground emission unrelated to CTB 80. 
We therefore have limited our spectral analysis of CTB 80 at $E> 0.5$~keV,
where we obtain $\sim$4800 background-subtracted counts. We rebinned the spectrum 
to contain at least 20 counts per energy channel, and fitted it with the X-ray 
emission spectral model of hot plasma in collisional
ionization equilibrium (APEC model in XSPEC). In our spectral model fits, we 
fixed the absorbing column at $N_{\rm H} = 3\times 10^{21}$~cm$^{-2}$, and the
abundances at solar \citep{anders89}.
The spectrum can be fitted with a two-component APEC model ($\chi^2/{\nu}$ = 117.9/106).  
The best-fit electron temperatures are $kT_{\rm hard} = 1.9^{+0.5}_{-0.3}$ keV and
$kT_{\rm soft} = 0.18{\pm}0.01$ keV. Their emission measures are  $EM_{\rm hard} = 
4.8 \times 10^{56}$~cm$^{-3}$ and 
$EM_{\rm soft} = 3.0\times 10^{57}$~cm$^{-3}$, respectively.
As in Appendix~\ref{sec:appendixB1}, 
we estimate the average electron density $\bar n_e$ of 
the soft component assuming $n_e \approx 1.2 n_{\rm H}$.
We also assume that the derived volume emission measure
is for a cylindrical volume with an angular radius $21'(=\sqrt{48'\times36'}/2)$ 
and a path length (along the LOS) corresponding to the physical diameter
of the spherical SNR ($2R = 64'$), 
Our estimated average electron density is $\bar n_e  \sim 0.08$ cm$^{-3}$. 
The mass of X-ray emitting gas inferred from the observed $EM$ is  
$M_X \approx 1.4m_{\rm H}\times EM/\bar n_e\sim 40$~\Msun.
Assuming that the interior of the SNR is filled with hot gas in pressure equilibrium, 
we estimate the thermal energy of the SNR to be 
$E_{\rm th} \approx 3 \bar n_e k_B T V_s  \sim 6 \times 10^{49} $ erg where 
$V_s$ is a spherical volume of radius $32'$ (=19 pc at 2 kpc).

\end{document}